\documentclass{emulateapj}
\usepackage{color,verbatim} 

\usepackage{epsfig}
\usepackage{subfigure}
\usepackage{graphicx}
\usepackage{amsmath}
\usepackage{natbib}
\newcommand{\pa}{\partial}
\newcommand{\mb}{\boldsymbol}

\shorttitle{Particle-Gas Dynamics in PPDs} \shortauthors{Bai \&
Stone}

\begin{document}


\title{Dynamics of Solids in the Midplane of Protoplanetary Disks: Implications
for Planetesimal Formation}


\author{Xue-Ning Bai \& James M. Stone}
\affil{Department of Astrophysical Sciences, Princeton University,
Princeton, NJ, 08544} \email{xbai@astro.princeton.edu,
jstone@astro.princeton.edu}




\begin{abstract}

We present local two-dimensional (2D) and three-dimensional (3D) hybrid numerical
simulations of particles and gas in the midplane of protoplanetary disks (PPDs)
using the Athena code. The particles are coupled to gas aerodynamically, with
particle-to-gas feedback included. Magnetorotational turbulence is ignored
as an approximation for the dead zone of PPDs, and we ignore particle self-gravity to
study the precursor of planetesimal formation. Our simulations include a wide
size distribution of particles, ranging from strongly coupled particles with dimensionless
stopping time $\tau_s\equiv\Omega t_{\rm stop}=10^{-4}$ to marginally coupled
ones with $\tau_s=1$ (where $\Omega$ is the orbital frequency, $t_{\rm stop}$ is the
particle friction time), and a wide range of solid abundances. Our main results are: 1.
Particles with $\tau_s\gtrsim10^{-2}$ actively participate in the streaming
instability, generate turbulence and maintain the height of the particle layer
before Kelvin-Helmholtz instability is triggered. 2. Strong particle clumping
as a consequence of the streaming instability occurs when a substantial fraction of
the solids are large ($\tau_s\gtrsim10^{-2}$) and when height-integrated solid to gas
mass ratio $Z$ is super-solar. We construct a toy model to offer an explanation.
3. The radial drift velocity is reduced relative to the conventional
Nakagawa-Sekiya-Hayashi (NSH) model, especially at high $Z$. Small particles may
drift outward. We derive a generalized NSH equilibrium solution for multiple particle
species which fits our results very well. 4. Collision velocity between particles with
$\tau_s\gtrsim10^{-2}$ is dominated by differential radial drift, and is strongly
reduced at larger Z. This is also captured by the multi-species NSH solution. Various
implications for planetesimal formation are discussed. In particular, we show there exist
two positive feedback loops with respect to the enrichment of local disk solid abundance
and grain growth. All these  effects promote planetesimal formation.

\end{abstract}



\section{Introduction}\label{sec:intro}

Planets are believed to be formed out of dust grains that collide and accrete
into larger and larger bodies in the gaseous protoplanetary disks (PPDs)
\citep{Safranov69,ChiangYoudin10}. The remarkable growth of dust into planets
covers 40 orders of magnitude in mass, and can be divided into three regimes.
At centimeter size or less, chemical bond and electrostatic forces allow
small dust grains to stick to each other to form larger aggregates
\citep{DominikTielens97,BlumWurm00,BlumWurm08}. At kilometer or larger sizes
(i.e., planetesimals and larger bodies), gravity is strong enough to retain
collision fragments, leading to the formation of planetary embryos/cores
\citep{WetherillStewart89,LissauerStewart93,KokuboIda98,Goldreich_etal04},
and ultimately to terrestrial and giant planets
\citep{Pollack_etal96,IdaLin04a,IdaLin04b,KenyonBromley06}. The intermediate
size range lies in the regime of planetesimal formation. This is probably the
least understood process in planet formation, largely because of solid growth
in this regime is subject to a bottleneck known as the ``meter size barrier".

In the intermediate size range, aerodynamic coupling between gas and solids is
important. The gaseous disk is partially supported by a radial pressure
gradient, and rotates at sub-Keplerian velocity, while solid bodies tend to
orbit at Keplerian velocity. Consequently, solid bodies feel a headwind and
drift radially inwards due to gas drag. The infall time scale is of the order
$10^2$ years for meter-sized bodies \citep{Weidenschilling77}, which poses
strong constraint on the timescale of planetesimal formation. Moreover, the
collision velocity between meter sized boulders and other bodies is large enough
to result in bouncing or fragmentation \citep{Guttler_etal10,Zsom_etal10},
rather than growth. To overcome the meter size barrier, collective effects that
form planetesimals out of meter sized or smaller bodies appear to be essential.
For example,  \citet{Cuzzi_etal01} proposed the turbulent concentration of
chondrule sized particulates by factors of up to $10^5$ by extrapolating
experimental results to high Reynolds numbers. It such dense regions, mutual
gravity of the particulates as a whole can overcome ram pressure and draw
them together to form planetesimals \citep{Cuzzi_etal08}, although the
intermittency in the turbulence might work against particle concentration
\citep{YoudinShu02}.

One favorable model of planetesimal formation involves gravitational
instability (GI) in the settled dust layer in the midplane of PPDs
\citep{Safranov69,GW73}. In the absence of turbulence in the disk, the dust
layer would become thinner and thinner until GI sets in and leads to formation
of planetesimals by gravitational collapse and fragmentation. However, as first
pointed out by \citet{Weidenschilling80}, turbulence generated by vertical
shear across the midplane dust layer (via the Kelvin-Helmholtz instability, hereafter
KHI) prevents dust grains from continuously settling well before GI is able to
operate. Based on the classical criterion for the onset of the KHI and solar
metalicity for height-integrated dust to gas mass ratio (hereafter, solid
abundance, denoted by Z), the maximum solid density in disk
midplane was found to be generally 1-2 orders of magnitude lower than the
Roche density for the onset of GI \citep{Sekiya98,YoudinShu02}
\footnote{The Roche density criterion for the onset of GI may not apply to
the dust sublayer due to the drag interaction between gas and solids, and
\citet{Youdin05a,Youdin05b} showed that GI can occur at lower densities with
smaller growth rate, although turbulent diffusion of solids is ignored in his
calculation.}. Inclusion of Coriolis force \citep{GomezOstriker05} as well as radial
shear \citep{Chiang08,Barranco09} do not alter the conclusion qualitatively. It
appears that increasing the local solid abundance by a factor of 2-10 times
solar is needed for this mechanism to operate\footnote{See also the most recent
results by \citet{Lee_etal10b} who studied the onset of KHI from more realistic
dust density profiles from dust settling.}. This factor may be achievable by
photoevaporation of gas \citep{ThroopBally05,AlexanderArmitage07}, and by
the radial variations of orbital drift speeds induced by gas drag
\citep{YoudinShu02,YoudinChiang04}.

An important ingredient of particle-gas interaction in the midplane solid
layer is the backreaction from particles to the gas. The momentum feedback
from solids to gas is responsible for KHI which tends to maintain a finite
thickness of the solid layer. When the solids are not too
strongly coupled to the gas, the backreaction leads to a powerful drag
instability \citep{GoodmanPindor00}, now termed the ``streaming
instability" (hereafter SI, \citealp{YoudinGoodman05}). The most remarkable
feature of the SI is that it very efficiently concentrates particles into dense clumps
\citep{YoudinJohansen07,JohansenYoudin07}, and enhances local particle
density by a factor of up to $10^3$. Such enhancement in particle density is
sufficient to trigger GI, and \citet{Johansen_etal07,Johansen_etal09} found in
their simulations that planetesimals form rapidly once self-gravity is turned
on. The sizes of the planetesimals formed in the simulations are about a few
hundreds kilometers, consistent with constraints deduced from observations of
asteroid and Kuiper belt objects that planetesimals are formed big
\citep{Morbidelli_etal09}. These results provide a very promising path for
forming planetesimals by SI followed by gravitational collapse.

Planetesimal formation is also affected by external turbulence in PPDs. The
typical mass accretion rate of $10^{-8\pm1}M_{\bigodot}\ $yr$^{-1}$
for T-Tauri stars \citep{Hartmann_etal98} indicates efficient angular momentum
transport in PPDs. Magnetic field seems certain to play a crucial role in the
transport process, most noticeably by the magnetorotational instability (MRI)
\citep{BH91,HB91}. The turbulence generated by MRI strongly affect the settling
of small dust grains \citep{FromangNelson09,Balsara_etal09,Tilley_etal10}, but
more interestingly, it promotes the concentration of decimeter to meter sized
bodies \citep{FromangNelson05,Johansen_etal06a,Johansen_etal07}. PPDs are,
however, only weakly ionized. The main ionization sources such as cosmic rays
and X-rays from the protostar only ionize the surface of the disk, making the
surface layers ``active" to MRI driven turbulence, while the midplane remains
poorly ionized and ``dead" \citep{Gammie96}. Accretion is therefore layered and
mainly proceeds in the active zone. Moreover, the presence of small dust grains
substantially increases disk resistivity and reduces the extent of the active layer
\citep{Sano_etal00,IlgnerNelson06,SalmeronWardle08,BaiGoodman09}. These
non-ideal MHD effects due to partial ionization and dust resistivity, as well as the
layered accretion structure in PPDs tremendously complicate the story of
planetesimal formation.

In this paper, we consider a {\it local} patch of PPDs and study the dynamics
of gas and solids in the disk midplane. We perform shearing box hybrid
simulations with both gas and particles using the Athena code \citep{AthenaTech}.
The implementation of the particle module and code tests are presented in
\citet{BaiStone10a}. The inclusion of backreaction from particles to gas allows
us to investigate both the SI and KHI simultaneously. The local model is necessary for
studying SI because the scale of particle clumping is much smaller than gas
scale height and requires at least $16$ cells to be properly resolved
\citep{BaiStone10a}. The self-gravity from particles is neglected. Although
self-gravity will ultimately play an important role in planetesimal formation,
our focus is its precursor: clumping of particles.
Neglecting self-gravity also has the advantage that our results can be easily
scaled to different disk parameters and have very broad applications (see
\S\ref{ssec:scaling}). We have also neglected the thermodynamics in our work,
which may affect the buoyancy of the gas, but the dynamics of the particles are
generally unaffected \citep{GaraudLin04}.

Our ultimate goal is to build the most realistic {\it local} model of PPDs possible,
including all of the non-ideal MHD effects as well as dust grains/solid bodies in
a self-consistent manner. In this paper, however, we focus on the dynamics in
the dead zone, and therefore can neglect MHD. This simplification is
justified in two ways. First, conductivity calculations have shown that the
inner part of PPDs ($r\lesssim10$AU) almost always contains a dead zone
\citep{BaiGoodman09,TurnerDrake09}.
Second, this approach separates the hydrodynamic effects (SI) from non-ideal
magnetohydrodynamic (MHD) effects, which sets the foundation for more
sophisticated work. In reality, the dynamics in the dead zone can be affected by
the turbulence in the active layer \citep{FlemingStone03}. For example, the gas
motion in the disk midplane may exhibit strong low-frequency (compared with
orbital frequency $\Omega$) vertical oscillations excited by the turbulence in the
upper layer, and no coherent anti-cyclic vortices are found \citep{OshiMacLow09}.
Its influence to the dynamics of the solids is not clear and is left for future investigations.


An important ingredient of our simulations is the size distribution of particles. A
wide size distribution of dust grains from micron to millimeter or centimeter
size in the PPDs is routinely deduced from the modeling of their spectral energy
distribution (SED) \citep{Chiang_etal01,Testi_etal03,DAlessio_etal06}.
Theoretical modeling of dust coagulation also result in a broad range of particle
sizes \citep{DullemondDominik05,Brauer_etal08a,Birnstiel_etal10}. In the most
recent work that incorporates up-to-date laboratory collision experiment
results \citep{Guttler_etal10,Zsom_etal10}, the particle size range that
dominates the total solid mass spans about 1-3 magnitude, typically from
sub-millimeter to decimeter range. We note that although
\citet{Johansen_etal07,Johansen_etal09} also considered a size distribution of
particles, their particle size is relatively large and the size range is narrow
(maximum particle size is 4 times the smallest). In this paper, we choose the
particle size range to span 1-3 orders of magnitude, and we assume uniform
particle mass distribution in logarithmic size bins.
Our choice of the particle size distribution roughly agrees with outcome of
coagulation model calculations and serves as a first approximation of reality. We
perform a parameter survey on particle size range and height-integrated particle
to gas mass ratio (or solid abundance) that cover a substantial fraction of parameter
space relevant to planetesimal formation. These simulations self-consistently
include the mutual interactions between gas and particles of all sizes
(extending the early analytical work by \citealp{Cuzzi_etal93} who assumed all
particles are passive), and will help
us better understand the environment and precursor of planetesimal formation.

We perform both two-dimensional (2D) and three-dimensional (3D) simulations,
where the 2D simulations are axisymmetric (i.e., in the radial-vertical plane).
We note that KHI is most prominent in the azimuthal-vertical plane
\citep{Johansen_etal06b}, although fully capturing KHI requires fully 3D
simulations including radial shear \citep{Chiang08,Barranco09,Lee_etal10}. On
the other hand, 2D simulation in the radial-vertical plane is sufficient to capture
SI \citep{YoudinGoodman05,JohansenYoudin07}. While 3D simulations are
necessary to capture all possible physical effects in the disk midplane layer,
we show in \S\ref{ssec:KHISI} that KHI is unlikely to be present in all our 3D
simulations, because the turbulence generated by SI is strong enough to
prevent the particles from further settling to trigger KHI\footnote{This is no
longer true if all particles are strongly coupled to gas, in which case the SI is much
weaker.}. Therefore, 2D simulations are also a valid approach to the problem,
and are much less time-consuming than the corresponding 3D runs. Moreover,
comparison between 2D and 3D simulations can be used for discerning
multi-dimension effects, and as a guidance for future studies.

This paper is organized as follows. In \S\ref{sec:setup}, we describe our
simulation method, model parameters and scaling relations. We also describe
the basic properties of the saturated state in all our simulations. We study
various aspects of our simulations in the subsequent four sections. In
\S\ref{sec:vertical} we discuss the vertical structure of the particle layer. In
particular, we address the question of what is the dominant process of the
midplane dynamics, KHI or SI? We further analyze which particles are actively
participating in the instabilities, and which particles behave only passively. In
\S\ref{sec:clumping} we study the conditions for forming dense clumps from
the SI, which preludes planetesimal formation. The composition and dynamics of
the dense clumps is also analyzed. \S\ref{sec:radial} deals with the radial
transport of particles, including both radial drift and radial diffusion. We study
particle collision velocities in \S\ref{sec:collision}. We conclude our paper in
\S\ref{sec:discussion} by summarizing our results and discussing various
implications for planetesimal formation. In particular, we summarize the
logical connections between various physical effects that may enhance
each other and promote planetesimal formation.


\section[]{Method and Simulations}\label{sec:setup}

\subsection[]{Formalism}

We consider local PPD models and formulate the equations of gas and solids
using the shearing sheet approximation \citep{GoldreichLyndenBell65}. We
choose a local reference frame located at a fiducial radius, corotating at
the Keplerian angular velocity $\Omega$. The dynamical equations are written
using Cartesian coordinates, with $\hat{\mb{x}},\hat{\mb{y}},\hat{\mb{z}}$
denoting unit vectors pointing to the radial, azimuthal and vertical
direction, where ${\mb\Omega}$ is along the $\hat{\mb{z}}$ direction. The gas
density and velocities are denoted by $\rho_g, {\mb u}$ in this non-inertial
frame. We include a distribution of particles coupled with gas via
aerodynamic drag, where the velocity of particle $i$ is denoted by ${\mb v}_i$.
The drag force is characterized by stopping time $t_{\rm stop}$, and equals
$({\mb u}-{\mb v})/t_{\rm stop}$ per unit particle mass.
Particles with different sizes have different
stopping times, labeled by subscript ``$k$". Back reaction from the particles
to gas is included, which is necessary for the study of KHI and SI. In this
non-inertial frame, the equations for the gas read
\begin{equation}\label{eq:gascont}
\frac{\pa\rho_g}{\pa t}+\nabla\cdot(\rho_g\mb{u})=0\ ,
\end{equation}
\begin{equation}
\begin{split}
\frac{\pa\rho_g\mb{u}}{\pa t}&+\nabla\cdot(\rho_g\mb{uu}+P\mb{I})
=\\
\rho_g&\bigg[2{\mb u}\times{\mb\Omega}+\Omega^2x\hat{\mb{x}}-\Omega^2z\hat{\mb{z}}+
\sum_{k}\epsilon_k\frac{\overline{\mb{v}}_k-\mb{u}}{t_{{\rm stop},k}}\bigg]\ .
\end{split}\label{eq:gasmotion}
\end{equation}
where the source terms include Coriolis force, radial tidal potential as well as
disk vertical gravity. The last term in the momentum equation represents the
backreaction (or momentum feedback) from particles to gas: $\epsilon_k$ and
$\overline{{\mb v}_k}$ denote the local mass density and velocity of particles of
type $k$. In this paper we neglect the effect of magnetic fields and focus on the
interaction between gas and solids in the dead zone of PPDs \citep{Gammie96}.
An isothermal equation of state for the gas is used throughout this paper, where
$P=\rho_gc_s^2$ and $c_s$ is the isothermal sound speed.

Similarly, the equation of motion for particle $i$ of type $k$ can be written as
\begin{equation}
\frac{d\mb{v}_i}{d
t}=-2\eta v_K\Omega\hat{\mb{x}}+2{\mb v}_i\times{\mb\Omega}+\Omega^2x_i\hat{\mb{x}}
-\Omega^2z_i\hat{\mb{z}}-\frac{\mb{v}_i-\mb{u}}{t_{{\rm stop},k}}\ .\label{eq:parmotion}
\end{equation}
In the above equation, we have added an inward force term $-2\eta v_K\Omega\hat{\mb{x}}$
to mimic the effect of an outward radial pressure gradient in the gas
\citep{BaiStone10a}, where $\eta v_K$ is the difference between gas velocity and
the Keplerian velocity in the absence of particles. This term will shift both gas and
particle azimuthal velocities by $\eta v_K$ relative to those in the real system. To
avoid confusion, we always use ${\mb u}$ and ${\mb v}$ to denote velocities that
corresponds to the real system (i.e., subtracting the azimuthal velocity component
from the simulation by $\eta v_K$).
Particle self-gravity is ignored as we focus on the dynamics in the midplane of the
PPD dead zone and precursor of the planetesimal formation.

In our simulations, we have applied an orbital advection algorithm for both
gas and particles \citep{StoneGardiner10,BaiStone10a}, and the actual velocities
used in the simulation are measured relative to the linearized Keplerian shear
flow: ${\mb u}'={\mb u}+(3/2)\Omega x\hat{\mb{y}}$ for gas flow, and
${\mb v}'_i={\mb v}_i+(3/2)\Omega x\hat{\mb{y}}$ for individual particles.

\subsection[]{Scaling Relations}\label{ssec:scaling}

Measuring velocities in units of the sound speed, time in units of $\Omega^{-1}$,
and length in units of the gas scale height $H_g\equiv c_s/\Omega$, the parameters
in the problem are reduced to the following:

\begin{enumerate}
\item The dimensionless particle stopping time $\tau_k\equiv\Omega t_{{\rm stop},k}$
for particle species $k$.

\item The solid abundance parameter $Z_k$ for each particle species, which
measures the height-integrated particle to gas mass ratio.

\item The parameter characterizing the strength of the radial pressure gradient
$\Pi\equiv\eta r/H_g=\eta v_K/c_s$.
\end{enumerate}

Below, we apply a disk model and provide the scaling relation between the
disk model parameters and these dimensionless parameters used in our simulation.

We adopt a generalized solar nebular model where the disk is vertically
isothermal and all the disk quantities have a power law dependence on the
radius \citep{YoudinShu02}
\begin{equation}
\begin{split}
 \Sigma_g&=1700f_gr_{\rm AU}^{-b}{\rm\ g\ cm}^{-2}\ ,\\
 T&=280f_Tr_{\rm AU}^{-c}{\rm \ K}\ ,\\
 M_*&=f_MM_{\bigodot}\ .
\end{split}\label{eq:nebula}
\end{equation}
where $\Sigma_g$ is the gas surface mass density, $T$ is the disk temperature,
$M_*$ is the mass of the central star, and $r_{\rm AU}\equiv r/1$AU. These
parameters fix the disk model. Although the global disk profile may not follow
the simple power law form, we can always approximate a local patch of the disk
in the above form, which is very general. In the standard minimum-mass solar
nebular (MMSN) model \citep{Hayashi81}, we have $b=3/2,\ c=1/2,\ f_T=f_g=f_M=1$.
The radial profiles of other physical quantities are
\begin{equation}
\begin{split}
\Omega&=2\pi f_M^{1/2}r_{\rm AU}^{-3/2}{\rm \ yr}^{-1}\ ,\\
v_K&=30f_M^{1/2}r_{\rm AU}^{-1/2}{\rm\ km\ s}^{-1}\ ,\\
c_s&=f_T^{1/2}r_{\rm AU}^{-c/2}{\rm\ km\ s}^{-1}\ ,\\
H_g&=3.4\times10^{-2}f_T^{1/2}f_M^{-1/2}r_{\rm AU}^{(3-c)/2}{\rm AU}\ .
\end{split}\label{eq:nebula}
\end{equation}
where in the calculation of the sound speed, we assume the mean molecular weight
$\mu=2.33$.

The background gas density profile is
\begin{equation}
\rho_{g,b}(r,z)=\frac{\Sigma_g}{\sqrt{2\pi}H_g}\exp(-z^2/2H_g^2)\ ,\label{eq:rhob}
\end{equation}
where subscript ``$b$" denotes ``background". Using this gas density and
sound speed, one can derive the radial pressure gradient in the gaseous
disk, thus obtain the amount of reduction $\eta v_K$ in the gas rotation
velocity. After some algebra, we can derive the pressure length scale
parameter
\begin{equation}
\begin{split}
\Pi\approx-\frac{1}{2}\frac{d\ln{P}}{d\ln{r}}\frac{c_s}{v_K}&=\bigg(\frac{3+2b+c}{4}-
\frac{3-c}{4}\frac{z^2}{H_g^2}\bigg)\frac{c_s}{v_K}\\
&\approx0.054f_T^{1/2}f_M^{-1/2}r_{\rm AU}^{1/4}\ .
\end{split}\label{eq:qparam}
\end{equation}
Note that $\Pi=\Pi(r,z)$ depends on both radius and height. Nevertheless,
in this paper, our simulation box is concentrated in the disk midplane
where $z\ll H_g$, therefore we can neglect the dependence of $\Pi$ on $z$.
In the last equation of the above formula, we have applied the power
law indices of the MMSN model. The dependence on disk temperature $f_T$,
stellar mass $f_M$ as well as disk radius $r$ is relatively weak. It is
worth mentioning that the dependence of $\Pi$ on disk mass is only through
the surface density profile parameter $b$, free from the scaling
parameter $f_g$. Therefore, the value $\Pi\approx0.05$ should apply to a
wide range of disk models.

Next we consider the scaling relations for the dimensionless stopping
time. Because the gas motion in PPDs is expected to be subsonic, the
relevant drag laws from the gas to the solids in PPDs are the Epstein
drag law \citep{Epstein24}, which applies when particle size is smaller than
the gas mean free path, and the Stokes drag law, which applies for larger bodies.
We assume all solid bodies have spherical shapes, then the stopping time in
these two regimes can be expressed as \citep{Weidenschilling77}
\begin{equation}
t_{\rm stop}=\begin{cases}
\dfrac{\rho_sa}{\rho_gc_s}\ , &
\text{$a<9\lambda_m/4$ (Epstein regime)}, \\
\dfrac{4\rho_sa^2}{9\rho_gc_s\lambda_m}\ , &
\text{$a>9\lambda_m/4$ (Stokes regime)}. \\
\end{cases}\label{eq:tstop}
\end{equation}
where $\rho_s\approx3$g cm$^{-3}$ and $a$ are the density and radius of
the solid body, $\lambda_m=(n_g\sigma)^{-1}=\mu m_H/\rho_g\sigma$ is the
mean free path of the gas, and $\sigma\approx2\times10^{-15}$cm$^2$ is
the molecular collision cross section \citep{ChapmanCowling70}. From the
above equations, we see that the particle stopping time depends linearly
on gas density in the Epstein regime. Nevertheless, the gas density can
be regarded as constant near the disk midplane where we study. Therefore,
in our local simulations, we can safely take $t_{\rm stop}$ as depending
on particle size $a$ only.

To better handle the relation between particle size and its corresponding
stopping time, we express the relation between $\tau_s\equiv\Omega t_{\rm stop}$
and $a$ by applying our disk model. The result is
\begin{equation}
\begin{split}
\tau_s=\max \bigg[
&4.4\times10^{-3}a_{\rm cm}f_g^{-1}r_{\rm AU}^b\ ,\\
&1.4\times10^{-3}a_{\rm cm}^2f_T^{-1/2}f_Mr_{\rm AU}^{(c-3)/2}\bigg]\ ,
\end{split}\label{eq:taus}
\end{equation}
where $a_{\rm cm}$ is the particle radius measure in centimeter. In the
MMSN model, at 1 AU, particles smaller than $3$cm are in the Epstein
regime. At larger radii, the Epstein regime applies to much larger
particles.

\subsection[]{Simulation Setup}

Fiducially, we consider the MMSN model at 1 AU, and set the
pressure length scale parameter $\Pi=0.05$. This parameter is kept fixed in all
our simulations. Instead of considering
a particle size distribution in radius, we consider the distribution in
$\tau_s$. Then one can easily translate it into particle radius given the
parameters of the disk model. We discretize a continuous particle size
distribution into a number of bins. Each bin covers half a dex in $\tau_s$ in the
logarithmic scale. For simplicity, we assume a uniform particle mass distribution
across the bins, that is, all the particle bins (or particle species) have equal
amount of mass. The parameters for the size distribution is therefore the minimum
and maximum dimensionless stopping time $\tau_{\rm min}$ and $\tau_{\rm max}$
(translated to $a_{\rm min}$ and $a_{\rm max}$ respectively). Physically, our
assumption means that most of the mass of the solids resides in the size range
between $a_{\rm min}$ and $a_{\rm max}$ and roughly follows a flat distribution
in logarithmic scale. To control the total particle mass, we use the total solid
abundance parameter
\begin{equation}
Z=\sum_{k=1}^{N_{\rm type}}Z_k\ ,\qquad {\rm with}\ Z_k=Z/N_{\rm type}\ ,\label{eq:massdist}
\end{equation}
where $N_{\rm type}$ is the
number of particle types (bins). Currently the best estimate of the solar metallicity
is about $0.015$ \citep{Lodders03}. A substantial fraction of the metal elements
may reside in dust grains and grow into larger bodies. In our simulations, we
consider three abundance values $Z=0.01, 0.02$ and $0.03$. This choice covers
a relatively wide range of disk metallicities. Moreover, because our simulation
focuses on a local patch in a PPD, the local abundance may not necessarily be
equal to the averaged value in the PPD.

As we explained in \S\ref{sec:intro}, we perform simulations in both 2D and 3D.
Our 2D simulations are in the $\hat{\mb x}$-$\hat{\mb z}$ plane (i.e. axisymmetric).
Details of the implementation and code tests of the particle-gas hybrid scheme
are given in \citet{BaiStone10a}. Our simulations use the standard shearing box
approach \citep{HGB95}, where the radial boundary condition is periodic with
azimuthal shear. Azimuthal boundary conditions are periodic. Vertical gravity is
included in our simulations, and we choose reflection boundary condition in the
$\hat{\mb z}$ direction, which is the same as that in \citet{Johansen_etal09}.
In general, we use 256 cells in the radial (and azimuthal, if applicable) direction.
Guided by \citet{BaiStone10a}, properly resolving the SI with $\tau_s=0.1$ requires
about $128$ cells per pressure length scale $\eta r$. With this required resolution,
our simulation box size is typically small, spanning  only about $2-4\eta r$. Such
small box size is also necessary to capture the typical wavelength of the KHI, if
present \citep{Johansen_etal06b}. In our simulations, we generally use
$N_p=65536$ particles per type for 2D simulations and $N_p=3145728$ particles
per type in 3D runs (in which cases $N_{\rm type}=7$). Larger $N_p$ are used
when $N_{\rm type}$ is smaller to keep the total number of particles similar in all
our simulations. Our choice of particle number guarantees at least one particle per
cell per particle type around the disk midplane, as required for numerical
convergence \citep{BaiStone10a}.

In our simulations, we set the initial particle density profile to be a Gaussian
centered on disk midplane with scale height $H_d=0.015H_g$ for all particle
types. The particle and gas velocities are computed from a multi-species
Nakagawa-Sekiya-Hayashi (NSH) equilibrium, where the classical
single-species NSH equilibrium \citep{NSH86} solution is generalized to
include multiple species of particles (see Appendix \ref{app:nsh}). Note that
different particles have different velocities, and the velocities of particles and
gas depend on $z$.

The choice of our simulation box size and boundary conditions in
the vertical direction merit further discussion. In the
simulations, gas-particle interaction in the disk midplane generates
turbulence and excites vertical motions in the gas. Ideally the
vertical box size should extend to a few $H_g$, similar to what is used for MRI
simulations (e.g., \citealp{SHGB96}), however, this would make 3D
simulations too expensive. We have conducted a series of tests in
2D with a single particle species $\tau_s=1$ using different vertical
box sizes and either reflecting or periodic boundary conditions.
In both cases, particles settle to the disk midplane with a spatial
distribution reminiscent of sinusoidal waves that slowly
drift in the radial direction. We find that the particle scale
height is more intermittent when using periodic boundary conditions.
Moreover, periodic boundary conditions appear to suppress 
asymmetric modes in the gas azimuthal velocity around the disk
midplane.  Using reflecting boundary conditions, we find essentially
no difference between the particle scale heights and clumping
properties obtained from different vertical box sizes once the box
height is much larger than the particle scale height, although it takes longer
for the system to reach a quasi-steady state when a larger vertical
box size is used. The drift velocities of the wave-like pattern of
particles do differ when different vertical box sizes
are used, but they are unlikely to affect the properties discussed
in \S\ref{sec:vertical} to \S\ref{sec:collision}. \footnote{Similar
tests have been performed using the Pencil Code with the same
conclusions (A. Johansen, private communication, 2009).}. Guided
by these results, as long as the vertical boundary of our simulation
box is well above the scale height of all particle species, one
should get converged results from the simulations.

Table \ref{tab:simulation} lists the parameters of all of our
simulations. Our runs are labeled using names with the form
R$xy$Z$z$-$n$D, where $x,y$ are integers corresponding to $\tau_{\rm
min}=10^{-x}$, and $\tau_{\rm max}=10^{-y}$, $z\equiv100Z$ represents
the solid abundance, and $n=2$ ($n=3$) denotes 2D (3D) simulations.
When referring to simulations with fixed $x$ and $y$ but all possible
values of $z$ and/or $n$, we omit the Z$z$, and/or the $nD$, parts
of the names.  We focus on two groups of runs.  In the first group,
the maximum particle stopping time is $\tau_{\rm max}=0.1$. We use
7 particle species to span three orders of magnitude in stopping
time (down to $\tau_{\rm min}=10^{-4}$) for the series of runs
labeled R41, while in the series labeled R21, we use three particle
species to span one order of magnitude in stopping time (down to
$\tau_{\rm min}=10^{-2}$).  In the second group of runs, the maximum
particle stopping time is $\tau_{\rm max}=1.0$, and the minimum
stopping time is chosen to be $\tau_{\rm min}=10^{-3}$ (R30) or
$0.1$ (R10).  In each series of runs (R41, R21, R30, R10), we perform
three 2D simulations with $Z=0.01, 0.02$ and $0.03$, and two 3D
simulations with $Z=0.01$ and $Z=0.03$.  Because of a smaller
$\tau_{\rm max}$ in the first group, higher resolution is needed
to resolve the SI.

\begin{table*}
\caption{Run parameters.}\label{tab:simulation}
\begin{center}
\begin{tabular}{ccccccccccc}\hline\hline
 Run & $Z/0.01\ ^1$ & $\tau_{\rm min}$ & $\tau_{\rm max}$ & $N_{\rm type}\ ^2$ & $L_x\times L_y\times L_z\ ^3$ &
 $N_x\times N_y\times N_z$ & $N_p\ ^4$ & $T_e\ (T_s)\ ^5$ \\\hline
 R41-2D & 1,2,3 & $10^{-4}$ & $10^{-1}$ & $7$ & $0.1\times-\times0.2$ & $256\times1\times512$ & $6.6\times10^{4}$ & $1500(1200)$ \\
 R41-3D & 1,3 & $10^{-4}$ & $10^{-1}$ & $7$ & $0.1\times0.1\times0.2$ & $256\times256\times512$ & $3.1\times10^{6}$ & $250(200)\ ^5$ \\
 R21-2D & 1,2,3 & $10^{-2}$ & $10^{-1}$ & $3$ & $0.1\times-\times0.2$ & $256\times1\times512$ & $9.8\times10^{4}$ & $900(600)$ \\
 R21-3D & 1,3 & $10^{-2}$ & $10^{-1}$ & $3$ & $0.1\times0.1\times0.2$ & $256\times256\times512$ & $6.3\times10^{6}$ & $250(200)\ ^5$ \\
 R30-2D & 1,2,3 & $10^{-3}$ & $1$       & $7$ & $0.2\times-\times0.3$ & $256\times1\times384$ & $6.6\times10^{4}$ & $1200(900)$ \\
 R30-3D & 1,3 & $10^{-3}$ & $1$     & $7$ & $0.2\times0.2\times0.3$ & $256\times256\times384$ & $3.1\times10^{6}$ & $450(300)$ \\
 R10-2D & 1,2,3 & $10^{-1}$ & $1$       & $3$ & $0.2\times-\times0.3$ & $256\times1\times384$ & $9.8\times10^{4}$ & $900(600)$ \\
 R10-3D & 1,3 & $10^{-1}$ & $1$     & $3$ & $0.2\times0.2\times0.3$ & $256\times256\times384$ & $6.3\times10^{6}$ & $450(300)\ ^5$ \\
\hline\hline
\end{tabular}
\end{center}

\qquad\qquad\qquad$^1$ Total particle to gas mass ratio, divided by 0.01.

\qquad\qquad\qquad$^2$ Number of particle species.

\qquad\qquad\qquad$^3$ Domain size, in unit of gas scale height $H_g=c_s/\Omega$. Note we have
fixed $\Pi=\eta r/H_g=0.05$.

\qquad\qquad\qquad$^4$ Number of particles per species in the simulation box.

\qquad\qquad\qquad$^5$ Total run time in unit of $\Omega^{-1}$. The number in the parenthesis
indicates the time of saturation. For R41-3D and R21-3D runs with Z=0.03, we have $T_e=280$ and
$T_s=240$. For run R10-3D, we have $T_e=500$ and $T_s=450$.
\end{table*}

\subsection[]{Simulation Runs and Saturation}

\begin{figure*}
    \centering
    \includegraphics[width=160mm,height=115mm]{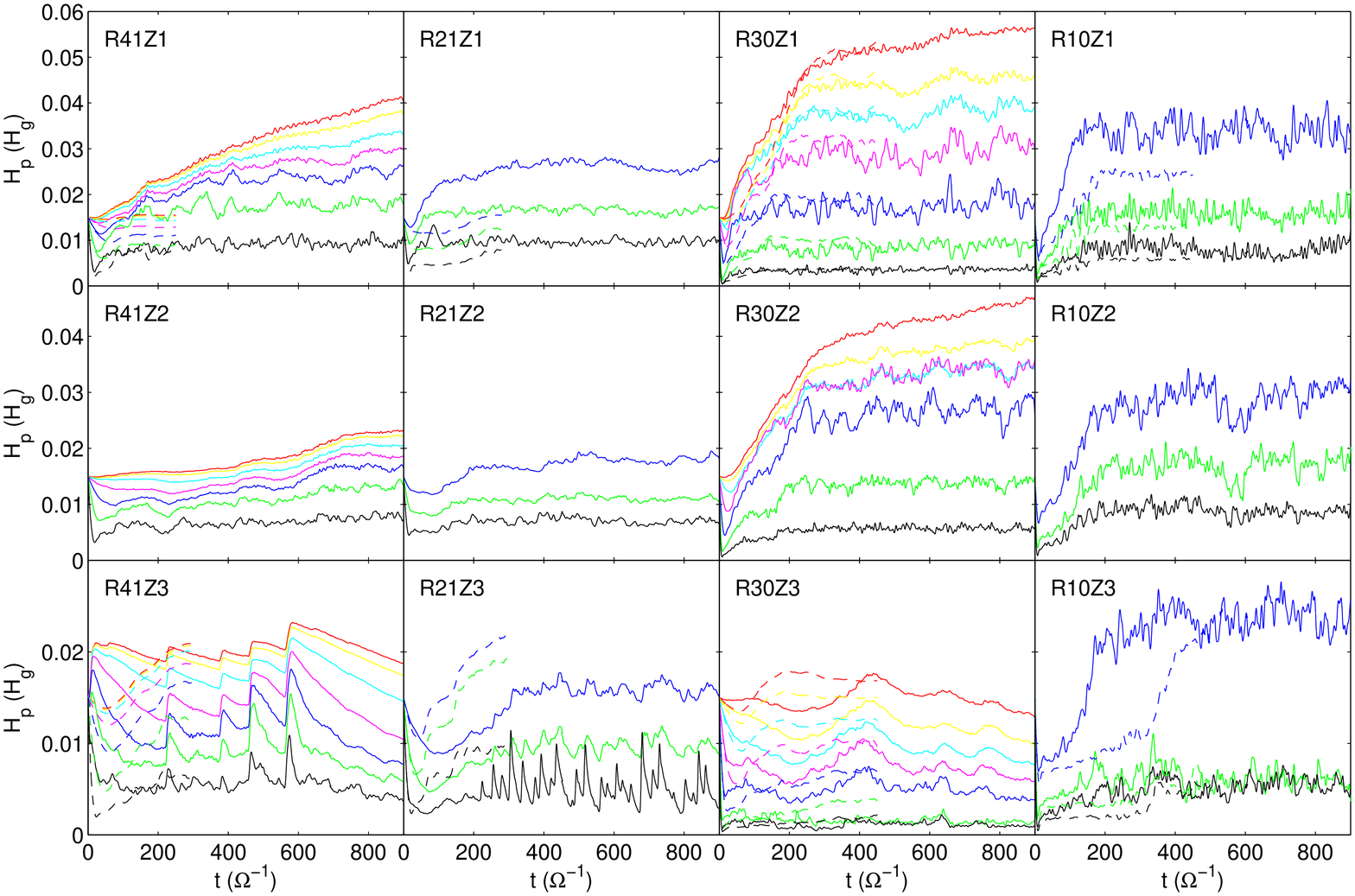}
  \caption{The time evolution of particle scale height $H_p$ for all simulations.
  Different colors represent different particle species, and particles with
  smaller $\tau_s$ have monotonically larger values of $H_p$ (see Table
  \ref{tab:simulation} for reference). Results from 2D simulations are plotted
  with solid curves, while dashed curves show 3D results. Note that
  we run 2D simulations much longer than those in 3D, and the vertical scale
  in the top, middle and bottom panels are different.}\label{fig:satdiag}
\end{figure*}

To determine when a saturated state is reached in each simulation,
we monitor the particle vertical scale height $H_{p,k}$ for each
particle species $k$, defined as the rms value of the $z$ coordinate
of all particles. Saturation occurs when particle settling and
turbulent diffusion are in balance, so that the scale height of all
particle species is steady. In Figure \ref{fig:satdiag} we show the
time evolution of the vertical scale height for each particle species
(marked by different colors) in all our runs. Solid and dashed
curves represent 2D and 3D simulations respectively. We see that
most of the 2D runs saturate within about 50 orbits\footnote{For
run R41Z1-2D, the diffusion time of the smallest particles with
$\tau_{s}=10^{-4}$ is very long and their $H_p$ still increases
after $1200\Omega^{-1}$. Nevertheless, the dynamics is dominated
by the largest particles with $\tau_{s}\gtrsim10^{-2}$, and the
scale heights of these particles has reached steady state.}. The
3D simulations are very time consuming, so we run them for shorter
periods. From Figure \ref{fig:satdiag}, all 3D runs saturate before
we terminate the simulations, although some just barely so.

In the last column of Table \ref{tab:simulation}, we
provide the time of saturation $T_s$ (in parentheses) for each
simulation. Unless otherwise stated, we will perform data analysis
in the time interval between the saturation time $T_s$ and the end
time of the simulation $T_e$. In the R41Z3-2D and R21Z3-2D runs,
there are sudden jumps in particle heights followed by settling, and
this process repeats over time quasi-periodically.  Averaging over
many cycles is required to reduce the influence of these intermittent
``bursts".  The vertical distribution of the smallest particles in
the R41-3D and R21-3D runs with $Z=0.03$ are not fully saturated
at the end of our simulations. Nonetheless, the scale heights of
the largest particles (which dominate the dynamics) in these runs
have reached steady state, therefore we consider them to be saturated.


\begin{figure*}
    \centering
    \includegraphics[width=180mm,height=180mm]{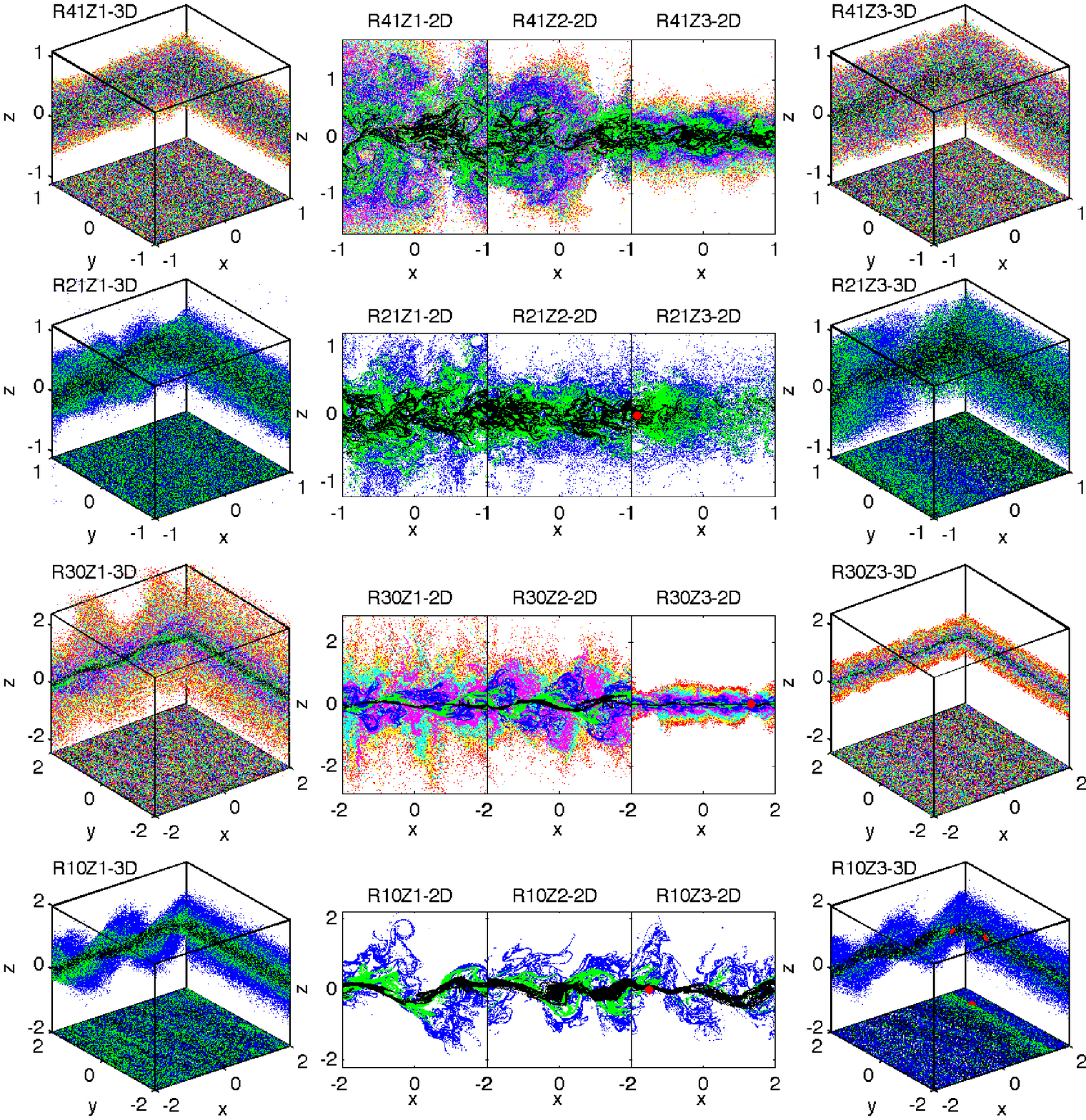}
  \caption{The distribution of particles at the end of all our simulations. For
  each 3D run (shown in the leftmost and rightmost panels), we show the
  projected positions of a subset of particles in three orthogonal directions,
  while each 2D run is shown in one panel in the center. Different particle
  species are marked with different colors, and the color coding is the same
  as that used in Figure \ref{fig:satdiag}. Large red dots in a few plots
  (corresponding to the simulation runs that exhibit strong particle clumping)
  indicate the densest point in the particle clump. The unit of length in all
  panels is $\eta r$. Note that the vertical size of our simulation box is larger
  than shown in this figure. }\label{fig:satpar}
\end{figure*}

Before presenting a detailed data analysis, we show the distribution
of particles at the end of our simulations in Figure \ref{fig:satpar}.
Results from 3D runs are shown by projecting particle positions in
three orthogonal directions. The number of particles plotted is
much less than the actual number of particles used in the simulation.
The trends in particle scale height evident in Figure \ref{fig:satdiag}
can be clearly seen: particles with small $\tau_s$ are diffused to
larger heights. Note that we overplot larger particles on top of
small particles, so that small particles near the midplane are less
visible. The SI is present in all the simulations, and we will discuss
various aspects of Figure \ref{fig:satpar} in the following sections.

\section[]{Vertical Structure of the Dusty Midplane Layer}\label{sec:vertical}


\subsection[]{Kelvin-Helmholtz Instability or Streaming Instability?}\label{ssec:KHISI}

The source of turbulence responsible for stirring up the particles can in
principle be due to both KHI and SI.
It is important to decipher which instability is the dominant process.
Generally speaking, the onset of SI requires the averaged
particle to gas mass ratio $\epsilon\gtrsim1$. The strength of the instability
decreases as the averaged particle size becomes smaller, and vanishes as
$\tau_s\rightarrow0$, for which the dust and gas behave as a single fluid. The
onset of KHI requires a steep vertical profile of gas azimuthal velocity, which
generally corresponds to larger dust to gas mass ratio at disk midplane. In our
simulations, a substantial fraction of the particles have a relatively large
stopping time with $\tau_s>10^{-2}$, and SI clearly plays an important role in
the generation of disk midplane turbulence. It remains to study whether KHI is
present and whether KHI is dynamically important.

The classical result on the onset of KHI in a vertically stratified disk is
based on the Richardson number criteria \citep{Chandra61}
\begin{equation}
Ri_{x,y}\equiv\frac{g}{\rho}\ \frac{(\pa\rho/\pa z)}
{(\pa u_{x,y}/\pa z)^2}\ ,\label{eq:RiDef}
\end{equation}
where we define the Richardson number from radial and azimuthal velocity shear,
as indicated by subscripts $x,y$. In the above equation, $g=\Omega^2z$ is the
vertical gravitational acceleration, and $\rho$ is the effective fluid
density (see discussion below). The Richardson number
measures the amount of work required to overturn the fluid (numerator) in
comparison to the amount of free energy available in the vertical shear
(denominator). For Cartesian flow with no rotation, the necessary
condition for instability is given by $Ri<Ri^{\rm crit}=1/4$. This criteria no
longer holds when rotation (Coriolis force) and radial shear (differential
rotation) are included, especially when the rotation frequency $\Omega$ is
comparable to the Brunt-V$\ddot{\rm a}$is$\ddot{\rm a}$l$\ddot{\rm a}$ frequency
of buoyant oscillations. Generally speaking, the Coriolis force distablizes the
fluid \citep{GomezOstriker05}, while radial shear acts to stablize the fluid.
\citet{Lee_etal10} found that $Ri^{\rm crit}$ is typically smaller than $1/4$ and is
roughly proportional to dust to gas mass ratio at disk midplane.
In this paper, we adopt the critical Richardson number to be $Ri^{\rm crit}=0.1$
as suggested by \citet{Chiang08}.

The Richardson number criterion is based on a single-fluid, in which
case $\rho$ simply represents fluid density. With the addition of perfectly
coupled dust, the dust-gas system behaves as a single fluid, where the dust
contributes to the mass but not the pressure of the fluid, thus $\rho=\rho_g+\rho_p$.
When particles are not perfectly coupled, the definition of $\rho$ becomes
somewhat ambiguous, but we expect $\rho_g<\rho<\rho_g+\rho_p$. Below we
provide a simple formula for $\rho$ in this regime that reduces the above two
limiting cases when $\rho_p=0$ and when $\tau_s\rightarrow0$.

In the absence of any turbulence and vertical gravity, the equilibrium state
between gas and dust (with fixed stopping time) is described by the NSH
solution \citep{NSH86}. In particular, the azimuthal gas velocity relative to
Keplerian velocity is given by
\begin{equation}\label{eq:uynsh}
u'_y=-\bigg[1-\frac{\epsilon(1+\epsilon)}{(1+\epsilon)^2+\tau_s^2}\bigg]
\eta v_K\ .
\end{equation}
where $\epsilon=\rho_p/\rho_g$, and prime means Keplerian
velocity is subtracted. For convenience, we define $\Delta u_y\equiv-u'_y$.
For perfectly coupled particles, $\tau_s=0$ and we find
$\rho_g+\rho_p=\rho_g\eta v_K/\Delta u_y$. For particles with finite stopping
time, $\Delta u_y$ becomes closer to $\eta v_K$, which reflects the fact that the
particle-gas coupling is weaker so that gas velocity shifts towards the dust-free
value. Therefore, $\Delta u_y$ can be regarded as an indicator of particle-gas
coupling. In this spirit, we define the effective gas density as
\begin{equation}
\rho^{\rm eff}\equiv\rho_g\frac{\eta v_K}{\Delta u_y}\ .\label{eq:rhoeff}
\end{equation}
It is trivial to check that in the limit $\epsilon\rightarrow0$,
$\rho^{\rm eff}\rightarrow\rho_g$, and $\rho^{\rm eff}\rightarrow\rho_g+\rho_p$
when $\tau_s\rightarrow0$. In the calculation of the Richardson number, we
substitute $\rho$ by $\rho^{\rm eff}$. Since $\rho_g$ is nearly constant over the
height of our simulation box, equation (\ref{eq:RiDef}) becomes
\begin{equation}
Ri_{x,y}=-\frac{\Omega^2z}{\Delta \overline{u}_y}\frac{(\pa\overline{u}_y/\pa z)}
{(\pa\overline{u}_{x,y}/\pa z)^2}\ ,\label{eq:RiDev}
\end{equation}
where the overbar means averaging over the horizontal plane. Note that $Ri$
depends on $z$.

\begin{figure*}
    \centering
    \includegraphics[width=180mm,height=60mm]{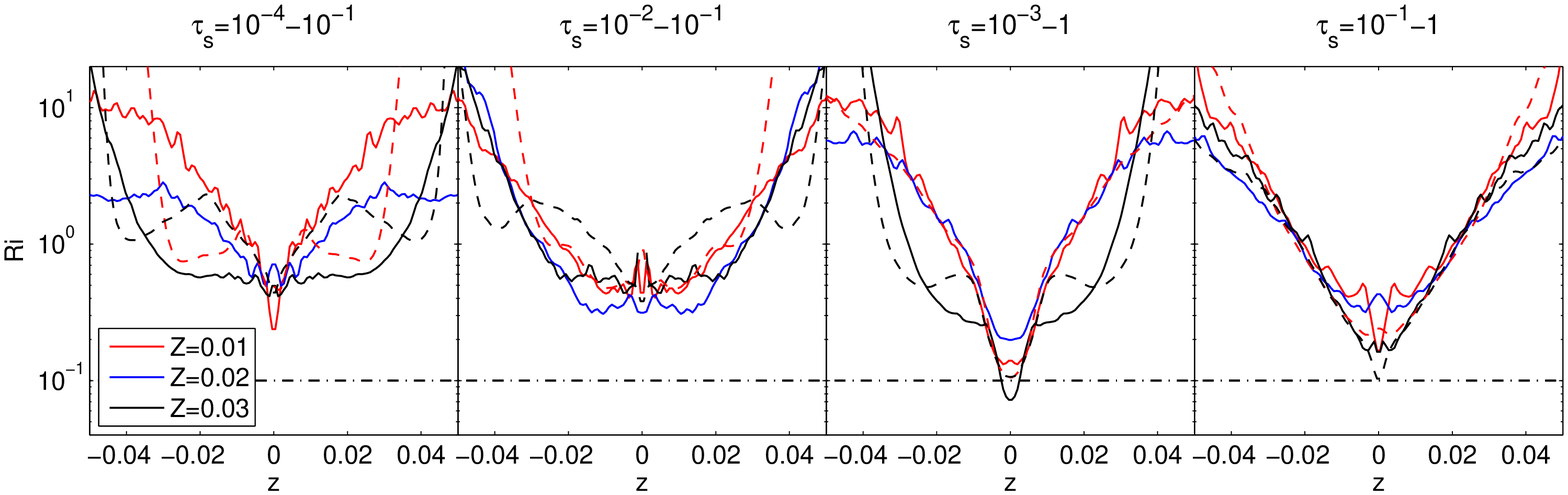}
  \caption{Profiles of the Richardson number computed from the gas azimuthal velocity
  from all simulations. Each panel shows the results from one series of runs. Red,
  blue and black curves label 2D simulations with $Z=0.01, 0.02$ and $0.03$
  respectively. The corresponding 3D simulations results are marked with the same
  colors but using dashed lines. Horizontal dash-dotted line marks the critical
  Richardson number $Ri=0.1$ adopted from \citet{Chiang08}.}\label{fig:richardson}
\end{figure*}

Before calculating the Richardson number profile from our simulations, we
first return to the spatial distribution of particles in Figure \ref{fig:satpar}.
In 2D simulations, we see that the distribution of particles around the disk
midplane is highly non-uniform, and exhibit wave patterns in the $x-z$ plane that
are almost stationary over time. Results from 3D simulations show very similar
features in the $x-z$ plane. In particular, in runs R30Z1-3D and R10Z1-3D, there
is a clear segregation of particles with different stopping times, and their wave
patterns have a phase shift relative to each other. However, in the $y-z$
plane, there is no coherent structure in the projected distribution of particles
in any of our 3D simulations. This contrasts with the expectations from the KHI,
where the particle layer kinks and breaks into clumps
\citep{Johansen_etal06b,Barranco09}. Based on this observation, we infer that
in our 3D simulations, KHI is not present in the azimuthal direction. Moreover,
in the $x-y$ plane, we see azimuthally elongated stripes of the large
particles (in black). This feature, together with the standing wave structure in the
$x-z$ plane, is most likely to be due to SI. KHI resulting from the vertical shear in
the gas radial velocity is another possibility, however, we have found that
$Ri_x$ is always larger than $Ri_y$ from our simulations, therefore the KHI is
unlikely to play a role in the simulations presented here.

In Figure \ref{fig:richardson} we show the Richardson number profile associated
from $u_y$ calculated from the saturated states of all our simulations. The
Richardson number is generally smallest in the disk midplane, and increases with
height. In almost all our 3D simulations (dashed curves), $Ri_y$ is greater than
the critical value (0.1), therefore, the dusty midplane layer is expected to be stable
against vertical shear, consistent with the spatial distribution of particles discussed
above. Given the fact that $Ri$ does not solely determine stability, this observation
does not entirely exclude the possibility that $Ri_y$ could be maintained by KHI.
However, it is important to note that KHI is suppressed in 2D. We see that $Ri_y$
from all our 2D simulations (solid curves) are generally close to their 3D
counterpart. This means that the SI itself is able to maintain $Ri$ above the critical
value, and suggests that the KHI is indeed absent in all our simulations.



The main reason that we do not observe KHI is that the turbulence
generated from the SI is strong enough to prevent particles from settling
sufficiently to trigger KHI. We note that the strength of the SI turbulence
decreases as the particle stopping time $\tau_s$ decreases (as
expected from the linear analysis of \citealp{YoudinGoodman05}, and
as confirmed by our numerical experiments). The turbulence in our
simulations is mainly generated from relatively large particles
with $\tau_s\gtrsim0.01$ (see also the next subsection). We have
not explored the regime where all particles are strongly coupled
to the gas.  However, in this regime, we expect the SI to be generated
on much smaller spatial scales with much lower amplitude, so that
the particles settle until the KHI is triggered. In this regime,
the dust-gas system behaves as a single fluid, where the dust
contributes to the mass density but not the pressure of the fluid.
This is the approach adopted by \citet{Chiang08}, \citet{Barranco09}
and \citet{Lee_etal10,Lee_etal10b} to study the KHI.



\subsection[]{Density Profile and Vertical Transport}\label{ssec:profile}

\begin{figure*}
    \centering
    \includegraphics[width=180mm,height=80mm]{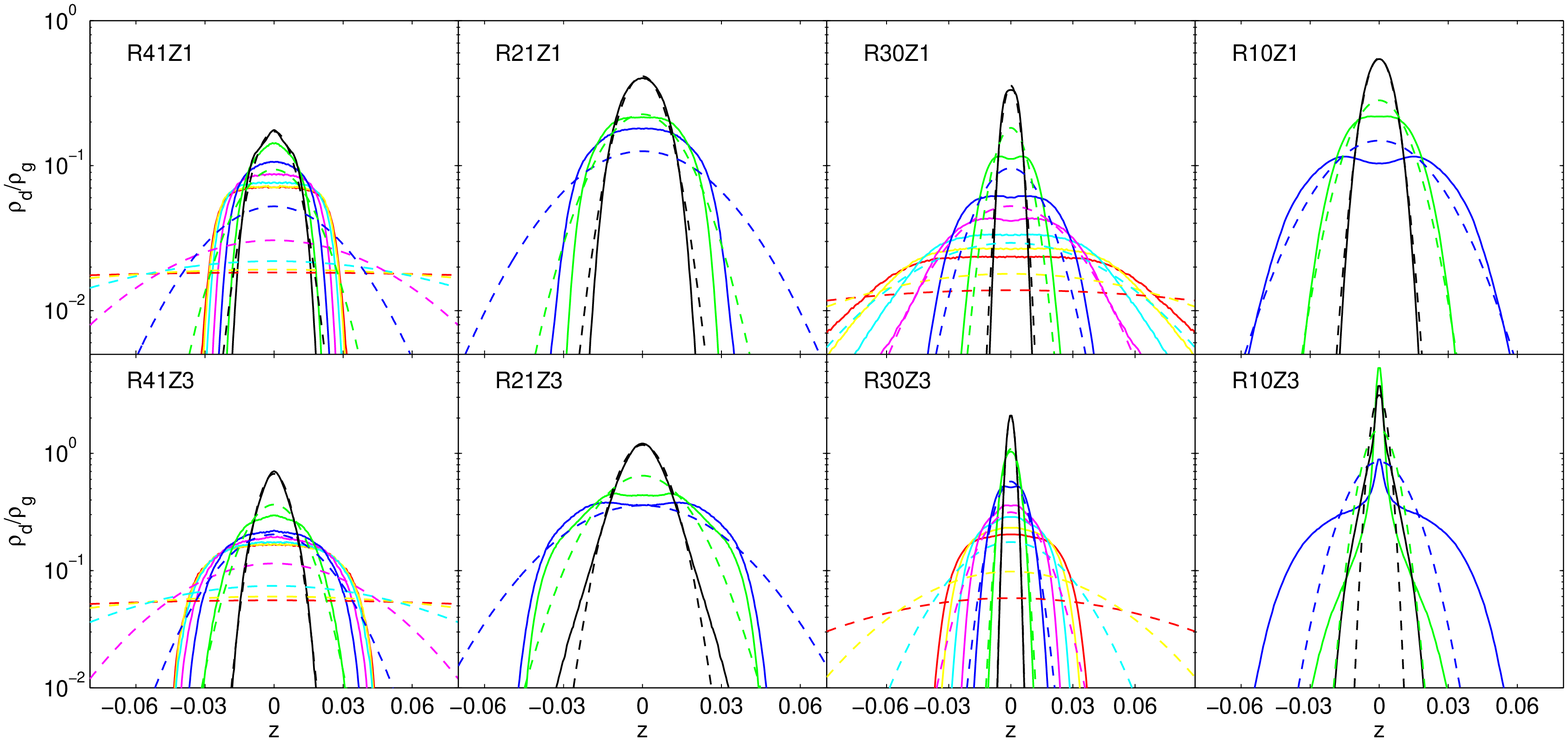}
  \caption{Vertical profiles of particle density from all our 3D simulations.
  Profiles of different particle types are labeled with different colors, with
  the same color scheme as in Figure \ref{fig:satdiag}, and particles with
  smaller stopping time have more extended profiles (see Table
  \ref{tab:simulation} for reference). Solid lines are time averaged particle density
  profiles from the saturated states of our runs, while dashed curves are
  model density profiles by assuming a constant turbulent diffusion coefficient
  fits to the density profile of particles with largest stopping time. See
  \S\ref{ssec:profile} for details.}\label{fig:profile3d}
\end{figure*}

Figure \ref{fig:profile3d} shows the vertical density profiles for particles of
different types from all our 3D simulations, calculated by binning the
particles into vertical grid cells and averaging over time after saturation.
Results from 2D simulations are generally similar.

\begin{table}
\caption{Vertical diffusion coefficient}\label{tab:vertdiff}
\begin{center}
\begin{tabular}{cccccc}\hline\hline
Run & $Z$  & $D_{g,z}$ (2D) & $D_{g,z} (3D)$ \\\hline
    & 0.01 & $2.05\times10^{-5}$ & $1.51\times10^{-5}$\\
R41 & 0.02 & $1.72\times10^{-5}$ & -\\
    & 0.03 & $5.09\times10^{-6}$ & $0.90\times10^{-5}$\\\hline
    & 0.01 & $2.12\times10^{-5}$ & $1.42\times10^{-5}$\\
R21 & 0.02 & $1.05\times10^{-5}$ & -\\
    & 0.03 & $1.97\times10^{-6}$ & $1.57\times10^{-5}$\\\hline
    & 0.01 & $5.14\times10^{-5}$ & $4.82\times10^{-5}$\\
R30 & 0.02 & $1.14\times10^{-4}$ & -\\
    & 0.03 & $6.28\times10^{-6}$ & $1.21\times10^{-5}$\\\hline
    & 0.01 & $1.91\times10^{-4}$ & $1.10\times10^{-4}$\\
R10 & 0.02 & $2.44\times10^{-4}$ & -\\
    & 0.03 & $5.63\times10^{-5}$ & $3.03\times10^{-5}$\\\hline\hline
\end{tabular}
\end{center}
The diffusion coefficients are measured in unit of $c_sH_g$.
\end{table}

The vertical density profile of particles is determined by the balance
between particle settling and turbulent diffusion. Unlike studies of
passive particles under the influence of homogeneous external turbulence
\citep{Cuzzi_etal93,YoudinLithwick07}, the turbulence from our simulations is
self-generated, and is non-homogeneous (strongest at the disk midplane).
To study the properties of turbulent diffusion, one approach would be to
assume some functional form for the vertical profile of the diffusion
coefficient $D_{g,z}(z)$, and fit the particle density profiles.
However, after several experiments we found it difficult to fit the density
profile of all particle species {\it simultaneously} with any simple functional
form of $D_{g,z}(z)$ \footnote{Part of the reason is that the Schmidt number
$Sc$, defined as the gas diffusivity divided by the particle diffusivity, is
uncertain.  In the limit $\tau_s\ll1$, one expects $Sc\rightarrow1$.
Even in this regime, we find the resulting profile $D_{g,z}(z)$ is not
described by any simple functional form that works for all our runs.}.
In fact, the wave patterns in the $x-z$ plane shown in Figure \ref{fig:satpar}
suggests that the classical turbulent diffusion scenario may be too simple.

Instead of fitting the vertical profile of the turbulent diffusion coefficient in the
gas, we pose the question in another way: What is the effective vertical
diffusion coefficient {\it at the disk midplane} for the particles that are driving
the turbulence? Since we have identified the SI as the source of the midplane
turbulence, one expects particles with relatively large stopping times to
drive the turbulence both from a theoretical point of view
\citep{YoudinGoodman05} and from non-stratified simulations of SI
(\citealp{JohansenYoudin07}, Bai \& Stone, unpublished). To address these
questions more quantitatively, we find the following approach particularly useful.

We fit the horizontally averaged vertical density profile of the largest
particles $\tau_s=\tau_{\rm max}$ in each simulation using the
classical picture of turbulent diffusion. Since these particles (as well as
particles with slightly smaller $\tau_s$) actively drive the disk turbulence, the
gas turbulent diffusion coefficient across this particle layer can be regarded as
constant. Therefore, the vertical density profile of these particles is expected
to be Gaussian, with scale height \citep{YoudinLithwick07}
\begin{equation}
H_p(\tau_s)=\sqrt{\frac{D_{g,z}(0)}{\Omega\tau_s}}
\sqrt{\frac{\tau_s+\tau_e}{\tau_s+\tau_e+\tau_s\tau_e^2}}\ ,\label{eq:Hp}
\end{equation}
where $\tau_e=\Omega t_{\rm eddy}$ is the turnover time of largest
eddies. The basic assumption behind this formula is stochastic turbulent forcing
on passive particles with the autocorrelation function of the turbulence
$P(t)=\exp{(-t/t_{\rm eddy})}/2\pi$, corresponding to a Kolmogorov spectrum. We
do not have much knowledge of $t_{\rm eddy}$ for SI turbulence, but expect
it to be comparable to the orbital time (the only time scale of the problem),
and take $\tau_e=1$. The exact value of $\tau_e$ does not matter much, since
it only gives an order unity correction to $H_p$.

By fitting the vertical density profile of the largest particles
with a Gaussian we obtain $D_{g,z}(0)$ for all simulations, and the
results are summarized in Table \ref{tab:vertdiff}. For 3D runs,
the results are also plotted in Figure \ref{fig:profile3d} as dashed
lines. We see that the vertical profiles of the largest particles
are well fitted with a Gaussian. In addition, we predict the vertical
density profile for other particle species, using equation (\ref{eq:Hp})
and assuming a diffusion coefficient which is constant with height.
Obviously, this will overpredict the scale heights for small particles,
since they respond to the turbulence passively. However, for particles
that actively participate in the instability, we expect their density
profile to be comparable to the predicted profile, since they are
driving turbulence to maintain $D_{g,z}$ close to $D_{g,z}(0)$ across
their scale heights.  In this way, we are able to identify the particle
species that are responsible for the disk turbulence (hereafter termed
as ``active" particles).

From the R41 runs, we see that active particles range from $\tau_s=0.1$
(for R41Z1) to $\tau_s\gtrsim0.01$ (R41Z3). Active particles for R21 runs
have $\tau_s\gtrsim0.03$. For R30 runs, particles with $\tau_s\gtrsim0.03$
are active, while for R10 runs, all particles are active. We see that
although there is a diversity in the size range of active particles, which depends
on both solid abundance and particle size distribution, the minimum size of
active particles for most of our runs is about $\tau_s=0.01-0.03$. For run R41Z1,
although we have identified somewhat larger $\tau_s$ values for active particles,
particles with $\tau_s=0.01-0.1$ must actively participate in the instability
because the abundance of $\tau_s=0.1$ particles alone is too small to trigger SI.

Next we study the midplane diffusion coefficient from our simulations. We
emphasize that the strength of the turbulence (hence $D_{g,z}$) is
self-regulated: the settling of particles continues until the turbulence they
generate is sufficient to stop the settling. To better interpret our results, we
construct a toy model describing the self-regulated turbulence. In this model, we
assume all particles are active, and that the particles are single-sized, with
fixed stopping time $\tau_s$. Since all particles are active, their vertical
density profile can be approximated by a Gaussian, so that the particle to gas
mass ratio at the disk midplane is given by $\epsilon=ZH_g/H_p$. The midplane
diffusion coefficient $D_{g,z}$ depends on both $\tau_s$ and $\epsilon$. For
simplicity, we parameterize the dependence as
\begin{equation}
D_{g,z}=\epsilon^\alpha f(\tau_s)H_g^2\Omega\ ,\label{eq:Dgparam}
\end{equation}
where $D$ is normalized to $H_g^2\Omega$, $f(\tau_s)$ is a coefficient that
incorporates the dependence of $D$ on $\tau_s$, and $\alpha$ is a power law
index that reflects the sensitivity of the dependence of $D$ on $\epsilon$. We
note that $D_{g,z}\rightarrow0$ at both $\epsilon\rightarrow0$ and
$\epsilon\rightarrow\infty$, therefore, we expect $\alpha>0$ when $\epsilon$ is
small and $\alpha<0$ for large $\epsilon$. Using equation (\ref{eq:Hp}) and
neglecting the second square root (which is order unity) on the right hand side,
we obtain
\begin{equation}
\begin{split}
H_p=&\bigg(\frac{f}{\tau_s}Z^\alpha\bigg)^{\frac{1}{2+\alpha}}\cdot H_g\ ,\\
D_{g,z}=&(fZ^\alpha)^{\frac{2}{2+\alpha}}
\tau_s^{\frac{\alpha}{2+\alpha}}\cdot H_g^2\Omega\ .\\
\end{split}\label{eq:toymodel}
\end{equation}
In the above equations, the dependence of particle scale height and diffusion
coefficient on $Z$ is reflected in the index $\alpha$. When $\alpha$ is positive,
increasing $Z$ leads to larger $H_p$ and larger $D_{g,z}$. When $\alpha$ is
negative, the situation reverses. Below, we apply this simple model to our
results. Since our simulations contain multiple particle species, we may take
$\epsilon$ to represent the contribution from all particle species participating in
the SI (i.e. with $\tau_s\gtrsim0.01$).

Our R30 and R10 runs show similar behavior between 2D and 3D simulations with
respect to
vertical diffusion properties. Increasing $Z$ from 0.01 to 0.02 produces stronger
turbulence, while further increasing $Z$ to 0.03 dramatically reduces $H_p$. This
corresponds to the transition from $\alpha>0$ to $\alpha<0$ at a threshold
$\epsilon$ (hence threshold $Z=Z_{\rm th}$). Beyond $Z_{\rm th}$, $H_p$
sensitively depends on $Z$ because the corresponding power law index
$\alpha/(2+\alpha)$ quickly drops to large negative values once $\alpha$ turns
negative. Consequently, a small increase in $Z$ results in strong particle settling
and greatly enhances midplane particle density. This result has
important implications for particle clumping discussed in the next section.

In our 2D R41 and R21 runs, we see that $D_{g,z}$ monotonically decreases with
$Z$, suggesting $\alpha<0$ for $Z\geq0.01$. Based on this result, we infer that the
strength of the SI for a particle size range $\tau_s=0.01-0.1$ is a decreasing function
of $\epsilon$ for $\epsilon\gtrsim0.5$. The 3D simulations give somewhat different results.
For both 3D R41 and R21 runs, $D_{g,z}$ slightly
increases with $Z$ at least in the range $Z\leq0.03$, indicating $\alpha\geq0$.
It is very likely that the threshold abundance $Z_{\rm th}$ is above $0.03$,
which is substantially larger than their 2D counterparts. We note that the behavior of
the SI turbulence for $\tau_s\lesssim0.1$ particles in 3D is different from that in 2D in
non-stratified simulations \citep{JohansenYoudin07}. Our results indicate that the
difference remains when vertical gravity is included, and 3D simulations are needed to
better catch the dynamics of small particles.


In our toy model, all of our ignorance on the dependence of
$D_{g,z}$ on $\tau_s$ is encapsulated in the unknown function $f(\tau_s)$. From Table
\ref{tab:vertdiff} we see that the R30 and R10 runs generally have larger $D_{g,z}$
than R41 and R21 runs. This result implies that turbulence generated from larger
particles $\tau_s\sim1$ is stronger than that from smaller particles, i.e.,
$f(\tau_s)$ is an increasing function of $\tau_s$ in this range, consistent with
results from non-stratified simulations \citep{JohansenYoudin07}.

In sum, we have identified that particles actively participating in SI
generally have stopping time $\tau_s\gtrsim0.01$. The strength of the
turbulence largely depends on the density of these active particles at disk
midplane. We find that the particle scale height (thus the turbulent diffusion
coefficient) strongly depends on solid abundance. Such strong dependence is
caused by a sharp drop in the strength of the turbulence with increasing particle to gas mass
ratio $\epsilon$ when $\epsilon$ is larger than a certain threshold value.





\section[]{Particle Concentration}\label{sec:clumping}

\subsection[]{Formation of Particle Clumps}\label{ssec:clumpform}

Probably the most interesting property of the SI is the concentration of particles.
The degree of particle concentration strongly depends on the mass distribution
of solids in PPDs. In our simulations, we normalize particle density to the
background gas density at the disk midplane $\rho_{g,b}(r,z=0)$. A useful
scale to measure particle concentration is the Roche density, above which the
particle clump can be considered as gravitationally bound \citep{BinneyTremaine08}
\begin{equation}
\rho_{\rm roche}(r)\approx\frac{3M_*}{r^3}=1.34\times10^3\frac{(f_Mf_T)^{1/2}}
{f_g}r_{\rm AU}^{\frac{2b-c-3}{2}}\rho_{g,b}(r,0)\ .\label{eq:roche}
\end{equation}
The normalized Roche density (relative to the background gas density at midplane)
scales as the square root of stellar mass and disk temperature, and is
inversely proportional to disk mass, meaning that the Roche density is easier to
reach for massive disks (with large $f_g$). In the MMSN model, the Roche
density is of the order $\rho_{\rm roche}=10^3\rho_{g,b}$, and only weakly
depends on $r$ as $r^{-1/4}$.

\begin{figure*}
    \centering
    \includegraphics[width=180mm,height=120mm]{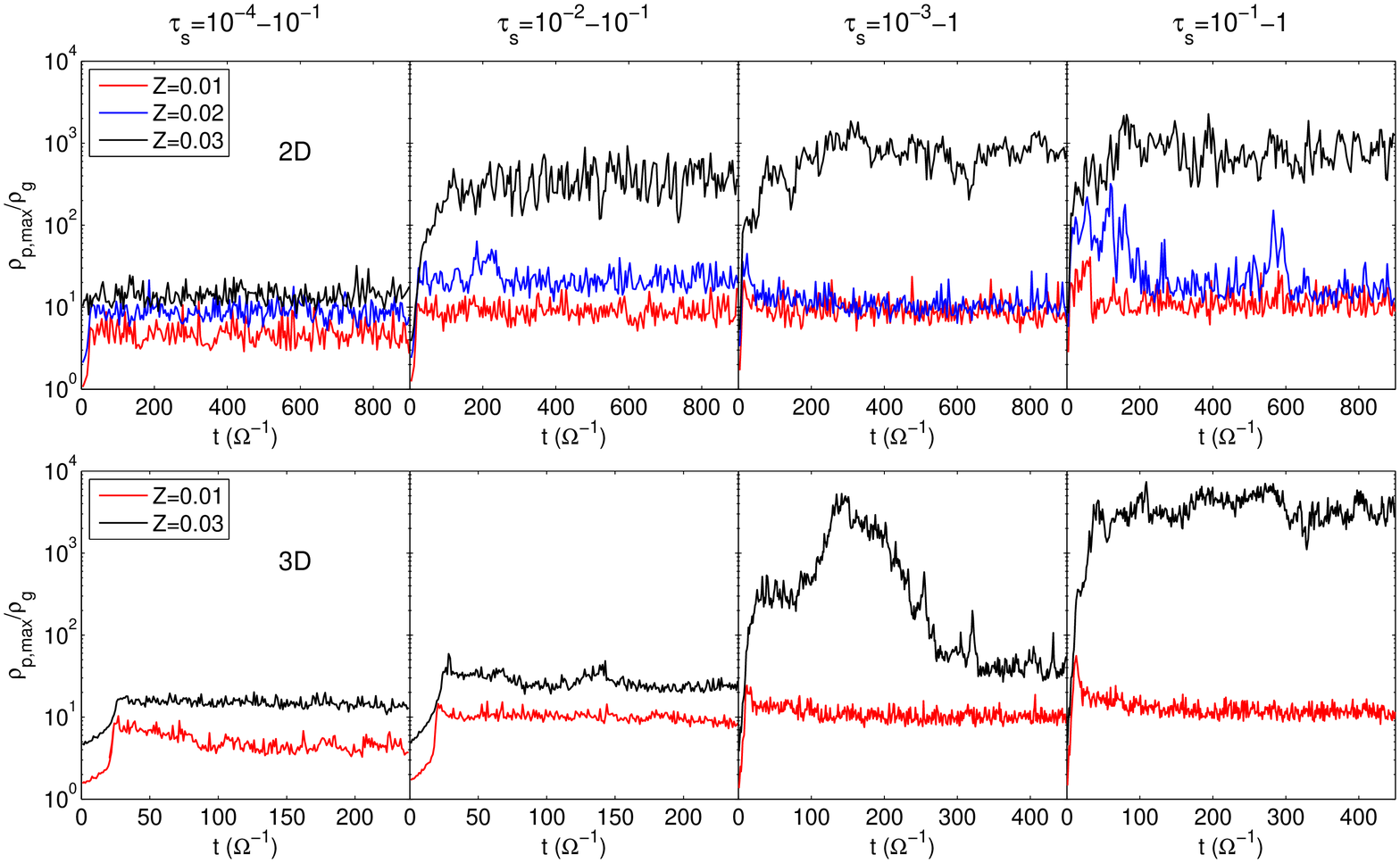}
  \caption{The evolution of maximum particle density for all our simulations.
  Each panel shows the results from one run series. In the upper panels,
  red, blue and black curves label 2D simulations with $Z=0.01, 0.02$ and $0.03$
  respectively. The 3D results are shown in the lower panels with $Z=0.01$
  and $0.03$ marked with red and black. The maximum density is normalized with
  respect to the background gas density at the disk midplane.}\label{fig:dparmax}
\end{figure*}

In Figure \ref{fig:dparmax} we show the time evolution of maximum particle
density $\rho_{p,{\rm max}}$ from all our simulations. We first look at
results from 2D simulations. For all the four run series,
$\rho_{p,{\rm max}}$ increases with solid abundance $Z$.
However, the dependence of $\rho_{p,{\rm max}}$ on $Z$ is highly non-linear. For
run series R21, R30 and R10, there is no significant clumping of particles for
$Z=0.01$ and $0.02$. However, significant clumping occurs at $Z=0.03$, with
maximum particle density reaching $10^3$ times the background gas density,
comparable to the Roche density (\ref{eq:roche}). This trend is consistent with
the results by \citet{Johansen_etal09} (see also the supplemental information in
\citealp{Johansen_etal07}), who considered particles with stopping time in the
range of $\tau_s=0.1-0.4$. As emphasized in the previous section, there is a sharp
enhancement of averaged midplane particle density with increasing $Z$ once
$Z$ exceeds some threshold value. This density enhancement further favors
strong concentration of particles by SI, which explains the trend we have
observed in Figure \ref{fig:dparmax}.

The particle clumping also depends on the particle size distribution. In the
R41 run series, where the majority of the particle mass resides in strongly
coupled particles $\tau_s<10^{-2}$, we see that there is no significant clumping
of particles up to $Z=0.03$. As noted in the previous section,
particles that effectively participate in SI are those with relatively large
stopping times $\tau_s\gtrsim10^{-2}$. These particles are also the ones that
actively participate in the clumping (see the next subsection). For R41 runs, the
abundance of these ``active" particles is much smaller than our R21, R30 and
R10 runs, which makes the critical (total) abundance for strong particle
clumping larger. In fact, we do observe strong clumping when we increase the
total abundance to $Z=0.05$. Based on the discussion above, we conclude that in
order for the SI to efficiently concentrate particles, the mass of the solids with
stopping time $\tau_s\gtrsim10^{-2}$ should exceed a critical value
$Z_{\rm crit}$. The results from 2D simulations suggest that
$\sum_{\tau_k\geqslant10^{-2}}Z_k \gtrsim Z_{\rm crit}\approx0.02$
is necessary for significant particle clumping\footnote{The value of the critical
metallicity also depends on the pressure gradient parameter $\Pi$
\citep{BaiStone10c}. A smaller value of $\Pi$ leads to smaller $Z_{\rm crit}$.}.

The 3D simulations show similar trends as in 2D, but the condition for strong particle
clumping is more stringent. Among the eight 3D runs, strong clumping
occurs only in run R10Z3-3D. The maximum density for all other runs remain small
in the saturated state ($\rho_{p,{\rm max}}\lesssim50\rho_{g,b}$). In particular, the
3D R21Z3 and R30Z3 runs do not show clumping as in their 2D counterparts, and
both of them have larger $D_{g,z}$. Since KHI is unlikely to be present in these
simulations, the different results between our 2D and 3D simulations should be
attributed to the different behavior of the SI in 2D and 3D. It appears that the formation
of dense particle clumps favors the mass distribution of particles to be dominated by
larger particles than in 2D, or larger values of $Z_{\rm crit}$ is needed.

Interestingly, in run R30Z3-3D, a very dense clump (actually a nearly axisymmetric
stripe) forms at about $t=150\Omega^{-1}$.
The composition of this (transient) clump is similar to its counterpart R30Z3-2D
(see next subsection). It lasts for about 10 orbital times and then is gradually
dissolved. Both the Richardson number profile and particle distribution disfavor the
presence of KHI during the process. Nor is there any significant vorticity
generation in the vicinity of the clump which might indicate KHI. By comparing with
Figure \ref{fig:satdiag}, we see that the period during which the clump is dissolved is
accompanied by an
increase of the height of relatively small particles with $\tau_s\lesssim0.1$. It is
likely that the formation of the transient clump is due to our unrealistic initial
condition\footnote{As small particles diffuse towards larger heights, the gas
azimuthal velocity at disk midplane is reduced, thus larger particles feel a stronger
headwind, enhancing the turbulence strength of the SI, which destroys the clumps.}.

The results we have obtained show a clear dichotomy on the particle concentration
properties. Specifically, the maximum density is either very small with
$\rho_{p,{\rm max}}\lesssim50\rho_{g,b}$, or very large with
$\rho_{p,{\rm max}}\gtrsim1000\rho_{g,b}$. Self-gravity becomes important when
the particle density approaches the Roche density (\ref{eq:roche}). This means
that for our simulations that do not show signature of strong clumping, adding
self-gravity will not change the picture qualitatively\footnote{Recent N-body
simulations by \citet{Michikoshi_etal10} show that gravitational collapse may occur
before Roche density is reached due to the drag force. This is unlikely to affect our
conclusion because in the non-clumping case $\rho_{p, {\rm max}}$ is usually more
than one order of magnitude smaller than the Roche density, and densest regions
are only transient.}. For simulations with strong
clumping, the maximum particle density is already comparable with the Roche
density, and in this case we expect the formation of a few planetesimals from the
simulations as in \citet{Johansen_etal09}.

Particle concentration properties are known to depend on numerical resolution. To
assess the validity of our results, we have also performed the same set of simulations
with half our standard resolution. We find the same dichotomy between strong
clumping and no clumping. The only exception is the R30Z3-3D run: it shows strong
particle clumping in the low-resolution run which does NOT dissolve as in our
standard resolution run. The reason is that the turbulence generated from the lower
resolution run is weaker, thus particles settle more which favors clumping. This
test justifies the necessity of conducting high resolution simulations. In the mean time,
it suggests that the critical abundance for particle clumping in this run may be only
slightly larger than $0.03$. Therefore, the particle clumping properties from 2D and
3D simulations is not dramatically different when $\tau_{\rm max}=1$.


\subsection[]{Properties of Dense Clumps}\label{ssec:clumpcomp}

\begin{figure*}
    \centering
    \includegraphics[width=160mm,height=60mm]{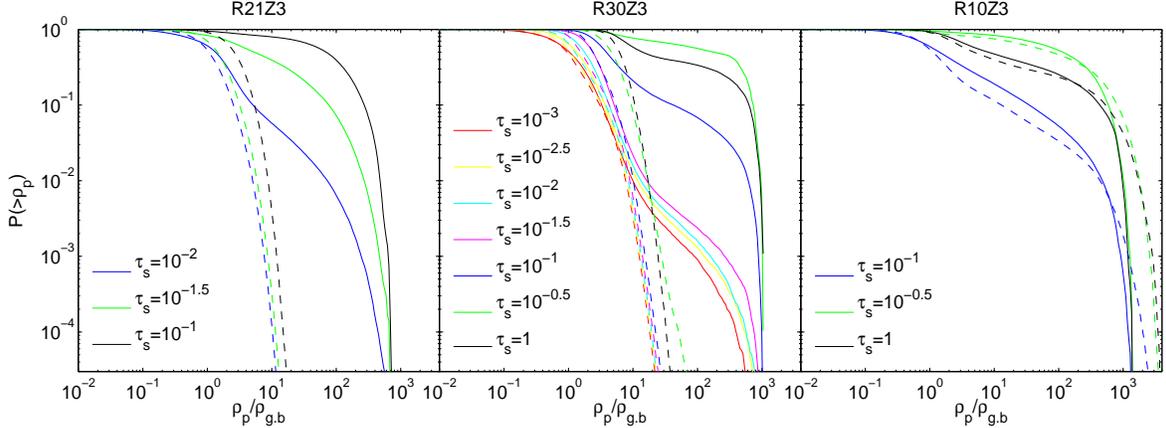}
  \caption{The cumulative probability distribution function (CPDF) of particle
  densities in the saturated state from the three simulations that exhibit
  strong particle clumping. Different
  colors mark the CPDF of different particle species, as labeled in the legend.
  Solid curves come from 2D simulations, while results from 3D simulations are
  denoted in dashed curves. Particle densities are normalized to background
  gas density at disk midplane.}\label{fig:dparpdf}
\end{figure*}

In this subsection we discuss more details of the three simulations that
exhibit strong particle clumping. First, we examine the composition of these
dense clumps by plotting the cumulative probability distribution function
(CPDF) of particle densities for different particle species $P(\rho_{p}>\rho)$.
The CPDF measures the probability of a particle residing in a region with total
particle density larger than $\rho$. In Figure \ref{fig:dparpdf}, we plot the
CPDFs of the three runs: R21Z3, R30Z3 and R10Z3. At relatively high
densities with $\rho_p\gtrsim10^2\rho_g$, we see that in all
three cases, the dense regions are composed of particles with the largest stopping
times. In run R21Z3-2D, the mass fraction of different particle species in the
dense clumps is increasing with particle stopping time $\tau_s$, and is
completely dominated by the largest particles $\tau_s=0.1$. In the case of
R30Z3-2D and R10Z3-2D, where the largest particles have $\tau_s=1.0$,
the composition of the clumps are dominated by the two largest particle species.
Contribution from other particle species to the clumps is almost negligible by mass.

For R21Z3 and R30Z3 runs, 3D simulations do not show particle clumping,
therefore, the resulting CPDFs differ substantially from those in 2D runs.
Nevertheless, these CPDFs provide typical examples for simulations
without clumping. The shapes of the CPDFs from different particles are very
similar, and curves for larger particles are located to the right of those for
smaller particles, consistent with the vertical stratification of particles. For run
R10Z3-3D, the particle clumping is stronger than the 2D case, and the densest
clumps are almost equally made of particles with $\tau_s=1$ and
$\tau_s=10^{-1/2}$.

\begin{figure*}
    \centering
    \includegraphics[width=180mm,height=110mm]{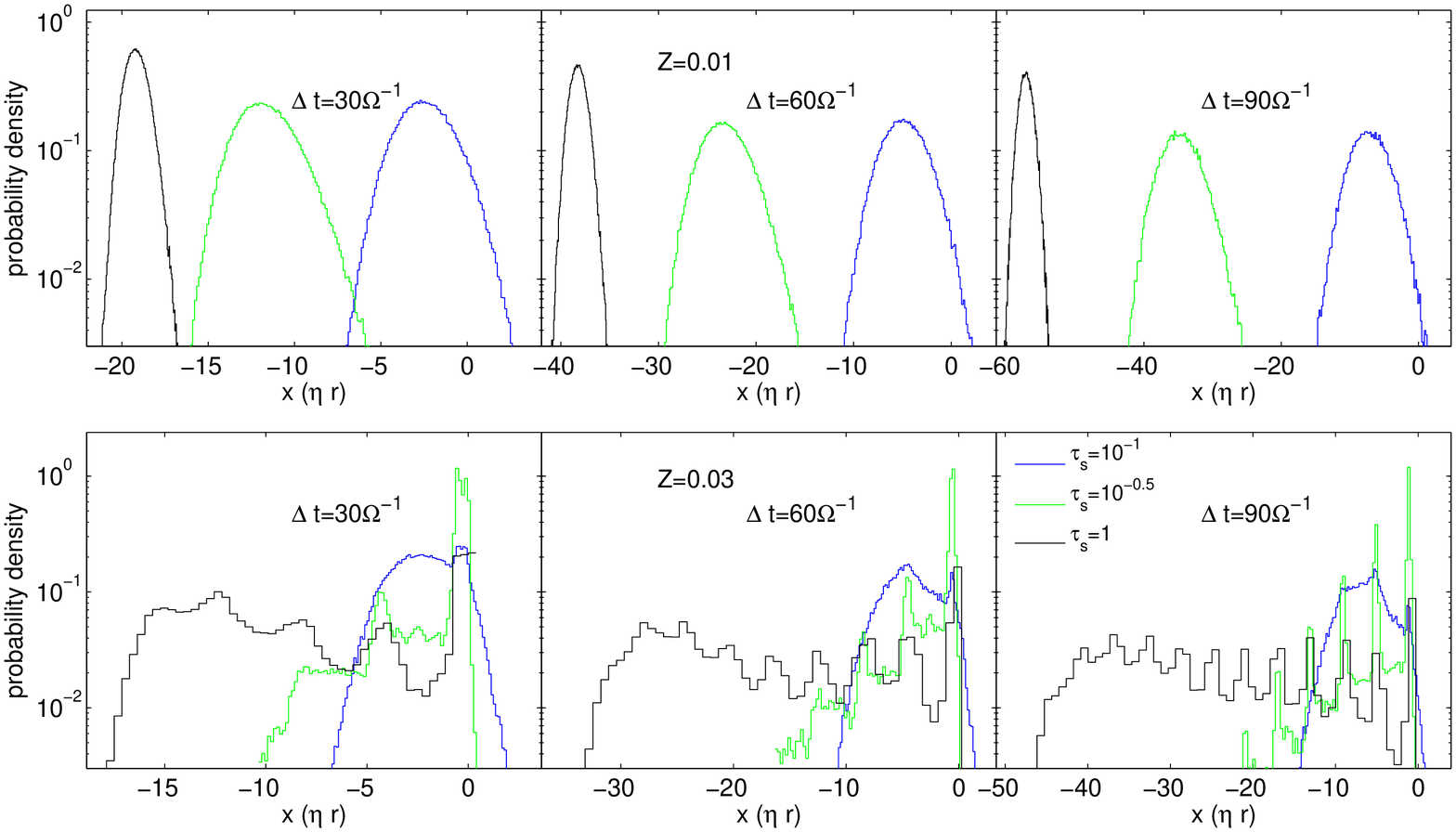}
  \caption{The probability distribution of the radial distance traveled by
  various types of particles after a given time interval $\Delta T$ from our
  runs R10Z1-3D (top) and R10Z3-3D (bottom). We choose
  $\Delta t=30$, $60$ and $90$ as shown in left, middle and right panels
  respectively (unit is $\Omega^{-1}$). Particles of different types are labeled
  by different colors (see the legend).}\label{fig:raddiffuse}
\end{figure*}

Next, we consider the motion of the dense clumps.
In Figure \ref{fig:satpar}, we mark the location of the densest point with a red
dot in runs with strong particle clumping. By monitoring the
location of the densest point with time, we find that it wanders slowly.
Another useful way of studying the dynamics of the clumps is by tracking the radial
trajectories $x_i(t)$ of a sample of particles. We relocate the particle
positions when they cross the radial boundaries of our simulation box so that
their trajectories are continuous. By tracing a large number of particles in the
saturated state of our runs, we obtain the distribution of $x(t+\Delta t)-x(t)$
for each particle species at time interval $\Delta t$. In Figure
\ref{fig:raddiffuse} we show the probability distribution of $x(t+\Delta t)-x(t)$
for a number particle species from our run R10-3D. When $Z=0.01$, no particle
clumping occurs. The distribution of $x(t+\Delta t)-x(t)$ is close to a Gaussian (or a
parabola in logarithmic scale) and the width increases with $\Delta t$, consistent
with undergoing a random walk. Meanwhile, the center of the distribution drifts
inward with time (see \S\ref{sec:radial} for more discussion). However, when
particle clumps are present, as in the $Z=0.03$ case, the shape of the
distribution deviates substantially from a Gaussian, especially for particles that
make up the clumps (the largest particles, shown in the blue and green curves).
For these clump-making particles, the width of particle distribution still increases
with $\Delta t$, as expected from turbulent diffusion, but a substantial fraction
of theses particles stay nearly stationary without drifting (near $x=0$), making
the resulting distribution more and more elongated with time. The leftmost
location of the particle distribution moves inward with time, and is set by the
radial drift velocity. More interestingly, we see almost evenly separated
multiple peaks in the distribution function. In fact, the separation between these
peaks equals the radial size of our simulation box. The physical picture
becomes clear that the clumps stop some of the particles from drifting radially,
and particles are kept in the clump for a few orbits or more before leaving for the
next clump. Similar behavior is observed for other runs with particle clumping.



\section[]{Radial Transport of Solid Particles}\label{sec:radial}

As expected from particle-gas equilibrium, particles experience head wind from
the gas and drift radially inward. Particles with different stopping times drift
at different velocities. At the same time, the instabilities generated at the disk
midplane diffuse the particles. These two processes transport particles radially
in PPDs, and is the subject of this section. In particular, we show that it is
important to study the radial transport of particles by considering particles of
all sizes simultaneously, rather than individually.

\subsection[]{Radial Drift Velocity}\label{ssec:drift}

\begin{figure*}
    \centering
    \includegraphics[width=180mm,height=135mm]{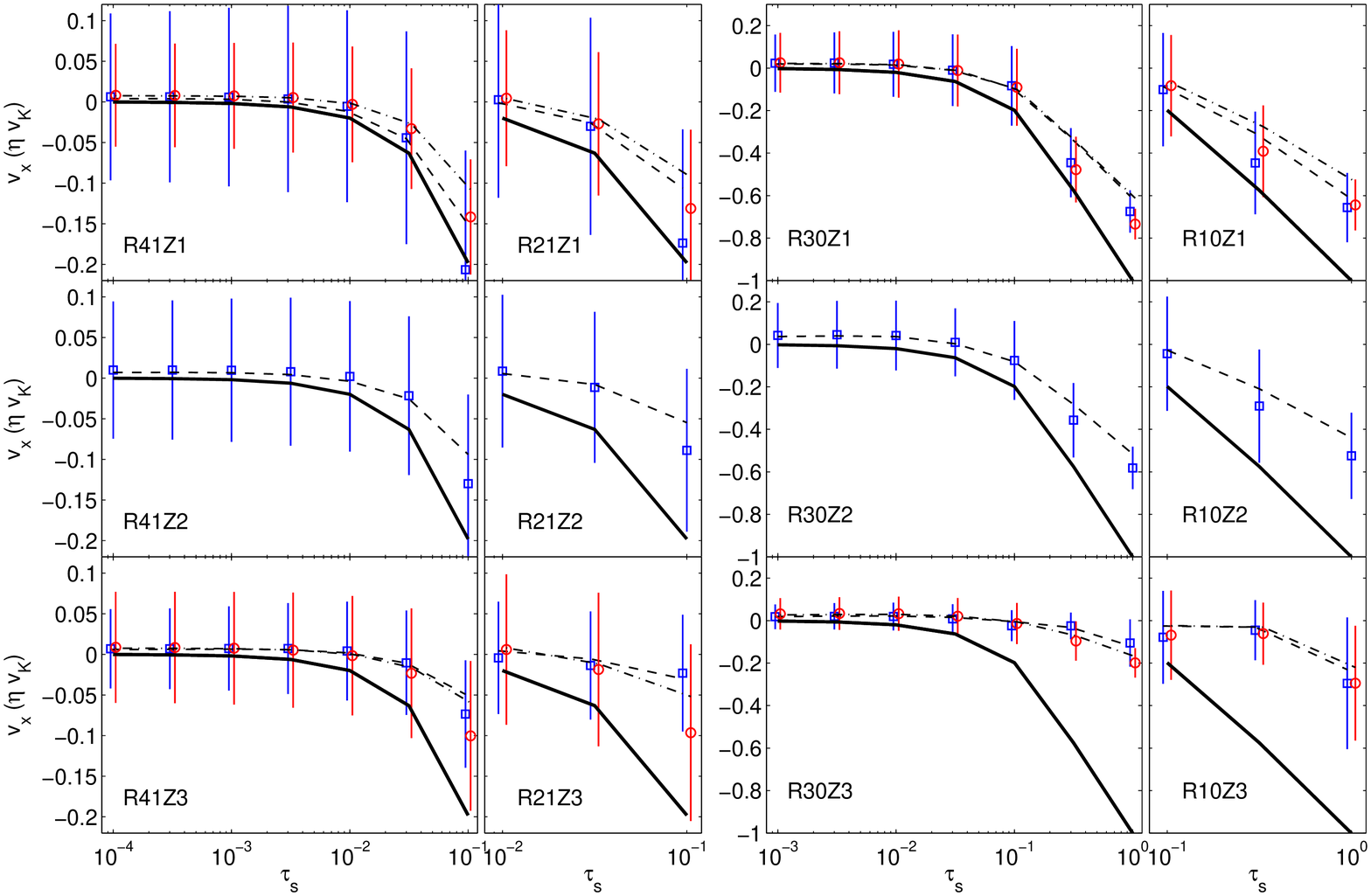}
  \caption{Radial drift velocities for different particle species from all our
  simulations. Blue squares: averaged particle radial drift velocity from 2D
  simulations, and vertical blue bars indicate its $1\sigma$ limits. Red circles:
  averaged particle radial drift velocity from 3D simulations, and vertical red
  curves show its $1\sigma$ limits. We have slightly shifted the symbols for
  2D and 3D runs in order to show the $1\sigma$ bars more clearly.
  Bold black solid line: radial drift velocity expected from
  single-species NSH equilibrium. Black dashed (dash-dotted) line: radial drift
  velocity expected from multi-species NSH equilibrium (see Appendix
  \ref{app:nsh}) for 2D (3D) simulations. All velocities are normalized
  to $\eta v_K$.}\label{fig:raddrift}
\end{figure*}

We calculate the averaged radial drift velocities for each particle species from
all our runs, and the results are shown in Figure \ref{fig:raddrift}.
The measured mean drift velocities are shown in squares (2D) and
circles (3D). We have also plotted the $1\sigma$ limits for particle drift
velocity based on the rms fluctuations, which are indicated in blue and red
vertical bars. In the figure, the velocities are normalized to $\eta v_K$. Clearly,
the radial drift velocity monotonically decreases with particle stopping time, and
the drift is fastest for marginally coupled particles.

The classical result on the radial drift of particles is the NSH equilibrium
solution \citep{NSH86}. It describes the equilibrium state between solids and
gas in unstratified (neglecting vertical gravity) Keplerian disks, where gas is
partially supported by radial pressure gradient. In the NSH equilibrium,
the drift speed is given by
\begin{equation}
v_x=-\frac{2\tau_s}{(1+\epsilon)^2+\tau_s^2}\eta v_K\ .\label{eq:singleNSH}
\end{equation}
We emphasize that the conventional NSH solution is obtained by considering a
single species of solids. Equation (\ref{eq:singleNSH}) does not simply generalize
to the case with multiple-species of particles by replacing $\epsilon$ to $\epsilon_k$
for each particle species $k$. In Appendix \ref{app:nsh} we provide the generalized
formula for multi-species NSH equilibrium, and the solution involves evaluation
of an inverse matrix of order $2N_{\rm type}$. It reflects the fact that although
different particle species do not interact directly with each other, they are
indirectly coupled via their interactions with gas.

In Figure \ref{fig:raddrift}, the bold solid lines show the expected radial drift
velocities from single-species NSH equilibrium. We see that there are large
deviations from the measured mean drift velocities, with two notable features.
First, for relatively large particles, the drift velocities are reduced from
single-species NSH values. The reduction is strongest
for runs with the largest $Z$. Second, the smallest particles drift
{\em outward}, rather than
inward as expected from the single-species NSH solution.

To calculate the expected radial drift velocity from a multi-species equilibrium,
we first use the particle density profiles extracted from \S\ref{ssec:profile} and
calculate the drift velocity in each vertical bin. The drift velocity is then weighted
by particle density in each bin to yield the mean drift velocity. The results are
plotted in dashed and dash-dotted lines (for 2D and 3D runs respectively)
in Figure \ref{fig:raddrift}. We see that these curves provide an excellent fit to
the measured mean radial drift velocities in all simulations. In fact, the two
features mentioned above are natural consequences of the multi-species
solution. Due to the sub-Keplerian motion of the gas, particle drag increases
gas angular momentum, leading to outward drift of gas. In the presence of both weakly
coupled and strongly coupled particles, the strongly coupled particles are tied
to the gas and therefore drift outward with the gas. Marginally coupled particles
still drift inward, but due to the influence of the smaller particles, these
particles feel a weaker headwind (i.e., the gas azimuthal velocity is closer to
the Keplerian value), resulting in a smaller drift
velocity compared with the single-species solution. With increasing $Z$, thus
higher midplane particle density, the gas becomes more entrained by
the solids, leading to stronger reduction of the drift velocity for large particles.

The residuals from the multi-species NSH solution fit to the measured mean drift
velocities are largest for particles with largest $\tau_s$, likely due
to their participation in SI, and/or clumping. In the non-stratified simulation of
\citet{JohansenYoudin07}, it was shown that in the saturated state of SI, the
radial drift velocity is either increased or decreased depending on run
parameters. In our simulations, these effects are secondary
compared with the multi-species effect. The measured drift velocities from 2D
(squares) and 3D (circles) simulations generally agree with each other. The
(small) differences can be attributed to the differences in the particle vertical
density profiles.

So far we have focused on the mean radial drift velocities. In the saturated
state of our simulations, the particle radial drift velocities follow a
distribution, due to the SI. We see in Figure \ref{fig:raddrift} that in most of the runs,
the fluctuation level is about $(0.05 - 0.15)\eta v_K$. This fact is closely
related to the radial diffusion of particles discussed in the next subsection. Based
this observation, we can estimate the particle radial diffusion coefficient to be
$D_x\sim(0.1\eta v_K)^2/\Omega\sim2.5\times10^{-5}c_sH_g$.

\subsection[]{Radial Diffusion}\label{ssec:raddiff}

\begin{figure*}
    \centering
    \includegraphics[width=150mm,height=130mm]{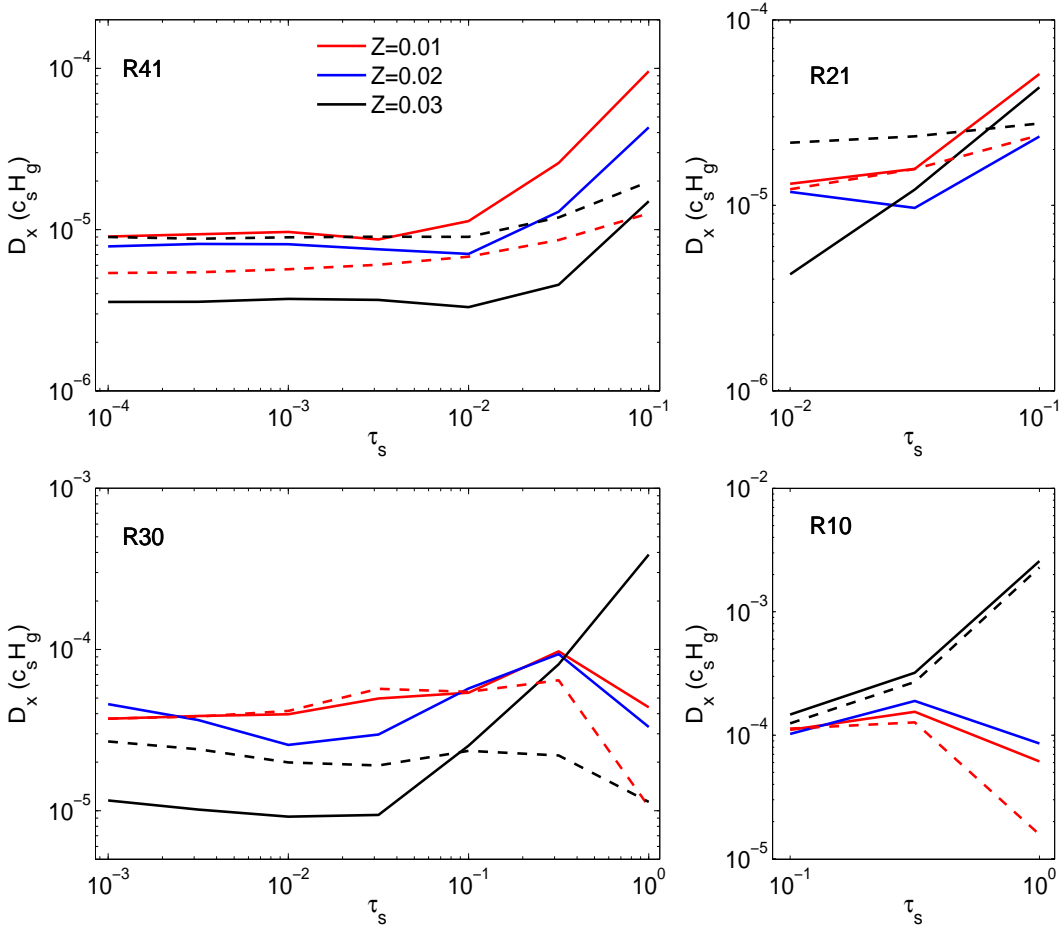}
  \caption{Radial diffusion coefficient for different particle species from all
  our simulations. Results from 2D and 3D simulations are shown in solid and
  dashed lines respectively. Red, blue and black curves represent different
  metallicities with $Z=0.01, 0.02$ and $0.03$ respectively. Diffusion coefficients
  are normalized by $c_sH_g$.}\label{fig:radmix}
\end{figure*}

The radial diffusion of particles is generally characterized by the radial diffusion
coefficient $D_x$. From our simulations, we can measure $D_x$ for different particle
species based on the random walk model of particle diffusion. We calculate the
distribution of shift in the particle radial position at various time intervals $\Delta t$ as
in Figure \ref{fig:raddiffuse}, and measure the width (rms) of the distribution
$\sigma$ as a function of $\Delta t$. The spreading due to a random walk
results in an Gaussian distribution, and $\sigma$ is related to the diffusion
coefficient by
\begin{equation}
D_x=\frac{1}{2}\frac{d\sigma_x^2}{dt}\ ,\label{eq:raddiff}
\end{equation}
For each particle species, we measure $\sigma_x^2$ for different $\Delta t$, and
fit the slope in of the $\sigma^2-\Delta t$ curve by linear regression. The
results are summarized in Figure \ref{fig:radmix}. The range of the radial
diffusion coefficient is consistent to within an order of magnitude of the estimate in the
last subsection based on the spread of radial drift velocities. It
is also comparable with the vertical diffusion coefficient at disk midplane
estimated in \S\ref{ssec:profile} (see Table \ref{tab:vertdiff}). Below we
discuss these results further.

First, the above procedure for measuring the diffusion coefficient does
not apply to runs that show strong particle clumping. As we see in Figure
\ref{fig:raddiffuse}, the distribution of $x(t+\Delta t)-x(t)$ deviates strongly
from a Gaussian due to the influence of the clumps. The measured
width of the distribution is about half the distance traveled by the fastest
drifting particles (those that are not confined in the clumps), and we observe that
$\sigma_x^2$ scales as $\Delta t^2$ rather than $\Delta t$ from our measurement.
Therefore, the measured $D_x$ from R21Z3-2D, R30Z3-2D and R10Z3 (both 2D
and 3D) runs for those clump making particles (or the largest two particle species
in the run) is not valid. In Figure \ref{fig:radmix}, we see the measured $D_x$ for
these particles have anomalously large values.
Such particles can reside in the disk for much longer than if there
were no clumping.

Next, we discuss diffusion of non-clumping particles. In each simulation
the measured $D_x$ generally approaches an asymptotic value for particles with
$\tau_s\lesssim10^{-2}$, but is different between different particle species for
particles with $\tau_s>10^{-2}$. This can be due to multiple reasons. First,
similar to the vertical diffusion of particles, the radial diffusion coefficient
also depends on the vertical position in the disk, and the radial diffusion in the disk
midplane is expected to be the strongest. Our measured $D_x$ can be considered as
a vertically averaged quantity. Therefore, $D_x$ is expected to be larger for
particles with larger $\tau_s$, since they stay closer to the midplane. This
trend is observed in runs R41 and R21. Second, different particles
react differently to the turbulence. In the case of Kolmogorov turbulence, the particle
diffusivity scales as $(1+\tau_s^2)^{-1}$ \citep{YoudinLithwick07}. This may be
responsible for the decrease of $D_x$ towards $\tau_s=1$ in R30 and R10 runs with
$Z=0.01$ and $0.02$. Thirdly, different particles participate in the SI in different
ways (i.e., actively or passively). The SI may strongly affect the transport
properties of the active particles, with the extreme example being the
clump-making particles discussed above. Despite the different values of $D_x$ for
different particle species, one may take the asymptotic value of $D_x$ as
measured from the smallest particles as characteristic of the radial diffusion
coefficient in the gas. These asymptotic values correlates with the vertical
diffusion coefficient well (see Table \ref{tab:vertdiff}).



To address the effectiveness of radial diffusion compared with radial drift, we denote
the mean radial drift velocity to be $v_r=\kappa\eta v_K$, and the diffusion coefficient
to be $D_x=(\beta\eta v_K)^2/\Omega$. After time $t$, the ratio
\begin{equation}
\zeta\equiv\frac{v_rt}{\sigma(t)}=\frac{\kappa}{\beta}\sqrt{\Omega t/2}\ ,
\end{equation}
reflects the relative importance between radial drift and turbulent diffusion, where
$\sigma(t)=\sqrt{2D_xt}$. Diffusion is important when $\zeta\lesssim1$. From Figure
\ref{fig:raddrift}, we see that for the largest particles, $\kappa\gtrsim0.1$. From
Figure \ref{fig:radmix}, we have $\beta\lesssim1$. Therefore, the
effect of radial diffusion of particles becomes negligible compared with radial
drift beyond $100$ orbital periods. Again, this discussion does not apply to the
situation when particle clumping is present, where large particles can be retained
in the clumps and some of them may survive the radial drift.


\section[]{Collision Velocities}\label{sec:collision}

The initial stage for planetesimal formation is the growth of solid
bodies by mutual collisions. The size distribution of particles in the PPDs
therefore depends on the outcome of two-body collisions, which further
depends on the properties of the colliding particles (e.g., size and porosity)
and collision velocity. Laboratory experiments show that at low collision
velocities ($\lesssim1$m$\cdot$s$^{-1}$), collisions generally lead to sticking
or bouncing. Larger collision velocities tend to result in fragmentation (see the
review by \citealp{BlumWurm08}). Nevertheless, sticking can also occur with
collision velocities up to $10-20$m$\cdot$s$^{-1}$ in some regimes (see Figure 11
of \citealp{Guttler_etal10}). The particle size distributions used in this paper
can be considered as a first approximation to the outcome of grain growth in
PPDs. In turn, we can measure the two-body collision velocity produced by the SI
from our simulations and investigate whether our selected particle size distribution
is consistent with the outcome of collisional coagulation.

\begin{figure*}
    \centering
    \includegraphics[width=180mm]{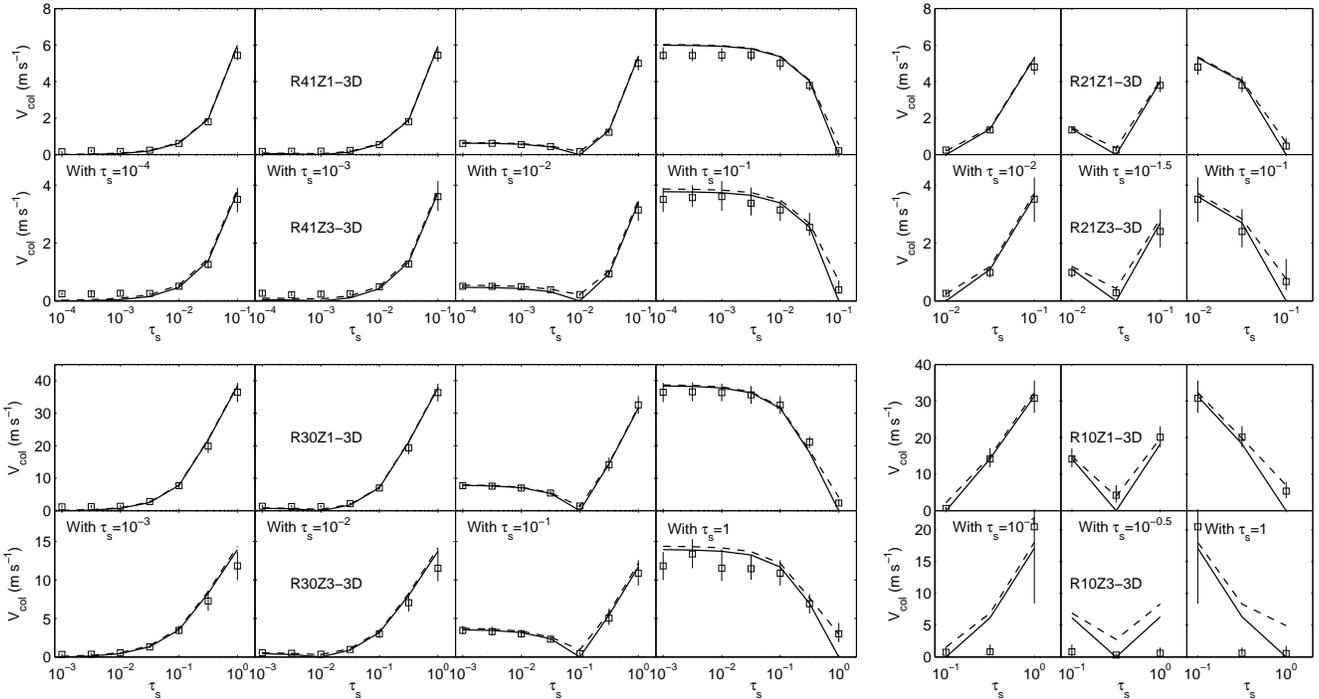}
  \caption{The median collision velocity for two-body particle collisions from
  all our 3D simulations. In each subplot, upper panels show results from
  metallicity $Z=0.01$ and bottom panels for $Z=0.03$. Plotted in each panel are
  median collision velocities between one particle with fixed $\tau_s$ and a
  second particle as a function of $\tau_s$ of the second particle. Squares:
  measured median collision velocities from the simulations. Vertical bars: the
  $1\sigma$ range of the collision velocities. Solid lines: expected collision
  velocity calculated from the radial drift velocities using multi-species NSH
  equilibrium (see Figure \ref{fig:raddrift}). Dashed lines: expected collision
  velocity from both radial drift and turbulence based on results from Figures
  \ref{fig:raddrift} and \ref{fig:radmix} (see text for details). The collision velocities
  scales linearly with the adopted sound speed, and we take $c_s=0.99$km s$^{-1}$
  appropriate for MMSN model at 1AU.}\label{fig:coll_mean}
\end{figure*}

We measure the relative speeds of all particle pairs within a distance
$\Delta r$ in the saturated state of our simulation snapshots. These velocities
form a representative sample of particle relative velocity distribution (RVD) in
the vicinity of a tracer particle. We assume that particles that collide with this tracer
particle would have the same RVD. The measured RVD depends somewhat
on the choice of $\Delta r$. In practice, we choose $\Delta r$ to be a
quarter of a cell size, in order to reduce the
(misrepresented) measured collision velocity between strongly coupled particles
(see Figure \ref{fig:coll_mean} and the discussion that follows), while maintaining
good statistics. To obtain the distribution of collision velocities with a tracer
particle, the RVD must be weighted by the relative velocity, since the collision
frequency is enhanced at larger relative velocities. The corresponding CPDFs
(similar to \S\ref{ssec:clumpcomp}) are shown and discussed in Appendix
\ref{app:collision}. In this context, it measures the probability of a particle that
undergoes collision with relative velocity greater than a given value. Particle
velocities are normalized to the gas sound speed $c_s$ in our simulations.
In all the results presented in this section, we adopt $c_s=0.99$km s$^{-1}$,
corresponding to the MMSN model at 1AU.

In order to visualize the particle collision velocities in a compact way, we
characterize the CPDFs by the median collision velocity (at $P=0.5$) and
its $1\sigma$ limits (at $P=0.68$ and $P=0.32$). In Figure \ref{fig:coll_mean}
we show the median collision velocities and $1\sigma$ limits
for various pairs of particle species from all our 3D simulations.
Results from 2D simulations are generally similar, and are not plotted. To interpret
these results, we consider two sources of the collision velocities:
radial drift and turbulence.

To calculate the contribution from radial drift, we evaluate the multi-species
NSH equilibrium in each vertical cell bin $j$ ($j=1,...,N_z$), from which we
obtain the relative radial drift velocity $(\Delta v_r)_{k_1,k_2}^j$ between each
pair of particle types $k_1,k_2$ in that bin. The relative velocity is further
weighted by collision frequency in that bin, proportional to
$(\Delta v_r)_{k_1,k_2}^j\epsilon_{k_1}^j \epsilon_{k_2}^j$. Integrating over all
the vertical bins, we obtain the expected collision velocity from radial drift,
which is shown as solid curves in Figure \ref{fig:coll_mean}. We see that with
the exception of run R10Z3-3D, these curves fit the median collision velocities
very well, meaning that relative radial drift is the dominant source of collision
velocities.

R10Z3-3D is the only 3D run that shows strong particle clumping, and the
measured median collision velocity is strongly reduced from our predictions.
This is clearly seen in the CPDF plot (see Figure \ref{fig:coll_pdf} in Appendix
\ref{app:collision}). However, in these simulations, the median collision velocity
no longer characterizes the overall collision velocities because the shapes of
the CPDFs are strongly deformed due to the clumping. In fact, there is still a
high-velocity tail in the CPDF of collision velocity, which reaches
values as high as $30$m s$^{-1}$. This tail is most likely caused by collisions
outside the clump, as indicated in Figure \ref{fig:coll_dpar},
and our predicted collision velocities should apply in these low density regions.

The relative radial drift velocity can not account for the collision velocity
between particles with the same stopping time (therefore all solid curves reach a
zero point in Figure \ref{fig:coll_mean}). To remedy this limitation, we further
consider the contribution from turbulence. So far turbulence induced particle
collision velocities has been studied theoretically only in the framework of passive
particles in uniform Kolmogorov turbulence \citep{Volk_etal80,Markiewicz_etal91},
and in MRI turbulence \citep{Carballido_etal08}. We consider the closed form
expression of turbulent collision velocities by \citet{OrmelCuzzi07}, which is based
on the Kolmogorov spectrum. Although these
assumptions do not quite apply in our simulations, we adopt this approach as an
approximate treatment of turbulence induced collision velocities. We
use their equation (16), and more specifically, we fix the turn over time for the
smallest eddy to be $t_\eta=0$, and take $y^*=t^*/t_{\rm stop}=1.6$ as an
approximation (where $t^*$ is the turn over time of the critical eddy with which
the particle in question is marginally coupled). The
turn over time of the largest eddy $t_L$, is considered as a fitting
parameter\footnote{In principle, $t_L$ is the same as $t_{\rm eddy}$ defined in
equation (\ref{eq:Hp}), where the latter is set to $\Omega^{-1}$ for simplicity.
Given the large uncertainties in this rough treatment of the turbulence induced
collision velocity calculation, we allow $t_L$ to vary.}. Because the strength of the
turbulence is vertically stratified, we take the averaged radial diffusion coefficient
$D_x$ from the smallest particles in each of our simulation run. The averaged
gas velocity $V_g$ is then related to $D_x$ by $D_x\simeq V_g^2t_L$.

In Figure \ref{fig:coll_mean}, we also show the contribution from turbulence
induced relative velocities as dashed curves. In order to fit the collision
velocity for pairs of large particles $\tau_s\gtrsim0.1$, we find
$\Omega t_L\simeq2-3$ for R41 and R21 runs, and $\Omega t_L\simeq4$ for R30 and
R10 runs. With this contribution, the collision velocity between the same types of
particles can be fit very well, and it also improves the fit to collision
velocities between particles with different types.

In our R41 and R31 runs, the predicted collision velocities almost reach zero for
collisions between particles with $\tau_s\lesssim10^{-3}$, since contributions
from both radial drift and turbulence rapidly decrease with stopping time. The
measured collision velocities are always larger than the predicted values, as seen
in the leftmost four panels of Figure \ref{fig:coll_mean}, and decrease towards a
small asymptotic value at smallest $\tau_s$.
We have experimented with choosing different $\Delta r$ in our calculations and
found that the asymptotic value roughly scales linearly with $\Delta r$ when
$\Delta r$ is less than grid size, because the gas velocity is not resolved at scales
less than a grid cell.

From Figure \ref{fig:coll_mean}, the median collision velocity is typically a
fraction of $\eta v_K$ ($\sim50$m s$^{-1}$ with our chosen scaling). Since the collision
velocity is dominated by the radial drift, and the radial drift is largest for
marginally coupled particles with $\tau_s\sim1$, we see that the collision
velocity is relatively small in the R41 and R21 runs (where $\tau_{\rm max}=0.1$),
typically smaller than $0.1\eta v_K$. The collision velocities from the R30 and R10
runs are much higher. Moreover, by comparing runs with the same particle size
distribution but different solid abundance, we see that the collision velocity is
reduced at larger $Z$. This is again due to the reduction of radial drift
velocity at larger $Z$ (see Figure \ref{fig:raddrift}). The typical value of the
collision velocity in our $Z=0.03$ runs are within $3$m s$^{-1}$ for R41 and R21
runs, and within $12$m s$^{-1}$ for R30 and R10 runs. Looking at Figure
11 of \citet{Guttler_etal10}, although collisions with relative velocity above
$1$m s$^{-1}$ are destructive in a number of situations, in other cases (e.g.,
when a porus particle hits a compact particle), particle growth is still
possible by mass transfer with collision velocities less than $10-20$m s$^{-1}$.
Detailed modeling of particle size evolution is beyond the scope
of this paper. Based on the results shown in Figure \ref{fig:coll_mean}, it is
possible for particle growth in all our R41 and R21 runs, as well as R30 and R10
runs with $Z=0.03$, meaning that the adopted particle size distribution in these
runs may be realizable. On the other hand, our R30 and R10 runs with $Z=0.01$
and $Z=0.02$ appear unlikely to be realized in nature, due to the destructive
collisions at velocities beyond $30$m s$^{-1}$. Combined with the results in
\S\ref{sec:clumping}, we conclude that larger solid abundance favors grain
growth in PPDs, which further promotes particle clumping.

\section[]{Discussion}\label{sec:discussion}

\subsection[]{Summary of Main Results}

The main purpose of this paper is to study the dynamics of solids and gas in the
midplane of PPDs using hybrid simulations. The solids and gas are coupled
aerodynamically, characterized by the dimensionless stopping time
$\tau_s=\Omega t_{\rm stop}$. We consider a wide size distribution of solids as
an approximation to the outcome of grain growth in PPDs, ranging from
sub-millimeter to meter size.
The key ingredient of our simulations is the inclusion of feedback from particles
to gas. Feedback is important when the local particle to gas mass ratio exceeds
order unity. Moreover, it is essential for the generation of SI and KHI.
In our simulations, we assume no external source of turbulence, as an
approximation for the dead zone of PPDs. Turbulence in the disk midplane is
generated self-consistently from the SI (driven by the radial
pressure gradient in the gas) and/or KHI (driven by vertical shear). Our simulations are local, since very high
numerical resolution is essential to resolve the SI and KHI. Self-gravity is ignored,
as we focus on the particle-gas dynamics before the formation
of planetesimals.

Our simulations are characterized by three sets of dimensionless parameters,
namely the particle size distribution $\tau_k$, solid abundance $Z$, and a
parameter $\Pi$ characterizing the radial pressure gradient. In this paper, we fix
$\Pi=0.05$, as appropriate for a wide range of disk model parameters (see
\S\ref{ssec:scaling}). The dependence of the particle clumping properties on
$\Pi$ is presented in a separate paper \citep{BaiStone10c}. We consider a flat
mass distribution in logarithmic bins
in $\tau_s$, and vary $Z$ from $0.01$ to $0.03$ (see Table \ref{tab:simulation}).
We conduct both 2D and 3D simulations, where 2D simulations are performed in the
radial-vertical plane in order for the SI to be actively generated. We run the
simulations for $40-200$ orbits and study the properties of the particles and gas
in the saturated state. The main results are summarized below.

\begin{enumerate}
\item SI plays the dominant role in the dynamics of PPD
    midplane when the largest solids have stopping times
    $\tau_s\gtrsim10^{-2}$. Particles with $\tau_s\gtrsim10^{-2}$ actively
    participate in SI, while smaller particles behave passively. KHI is not
    observed in all our simulations, which suggests that it may be important
    only when all particles have $\tau_s\lesssim10^{-2}$.

\item The strength of the turbulence generated by the SI and the scale height of
    the particle layer are self-regulated. There exists some threshold solid
    abundance, above which increasing $Z$ will result in weaker turbulence,
    which promotes particle settling, leading to rapid drop of the thickness of
    the particle layer and strong particle clumping.

\item SI can concentrate particles into dense clumps with solid density
    exceeding the Roche density, which acts as the prelude of planetesimal formation.
    The particle clumping generally requires the presence of relatively large
    particles with $\tau_s\gtrsim10^{-2}$. It also sensitively depends on solid
    abundance, in favor of super-solar metallicity.

\item The dense particle clumps are mostly made of the largest particles
    with size range spanning less than one order of magnitude. These
    particles are trapped in the clumps for several orbital times before leaving
    the clumps, providing a way for large particles to survive radial drift.

\item The mean radial drift velocity for each particle species
    agrees well with a multi-species NSH equilibrium solution
    (see Appendix A). Strongly coupled particles drift outward, and
    the radial drift velocity for particles with larger $\tau_s$ is strongly reduced
    relative to the conventional single-species NSH value, especially at large $Z$.
    This can increase the lifetime of the largest particles by a factor of a few.

\item Turbulence generated by the SI leads to radial diffusion of particles,
    but the diffusion is slow and its effect is negligible compared with
    radial drift after about $100$ orbital periods (for the largest particles).
    Particle clumping effectively enhances radial diffusion by retaining a
    fraction of large particles in the clumps.

\item Mutual collision velocity between $\tau_s\gtrsim10^{-2}$ particles is
    dominated by the difference in their radial drift velocities, and agrees
    well with calculations using the multi-species NSH equilibrium. The
    collision velocity is strongly reduced towards large disk metallicity
    relative to predictions from single-species NSH solution. Collision
    velocity induced by SI turbulence is only secondary.

\end{enumerate}

\subsection[]{Implications for Planetesimal Formation}

\begin{figure}
    \centering
    \includegraphics[width=80mm,height=90mm]{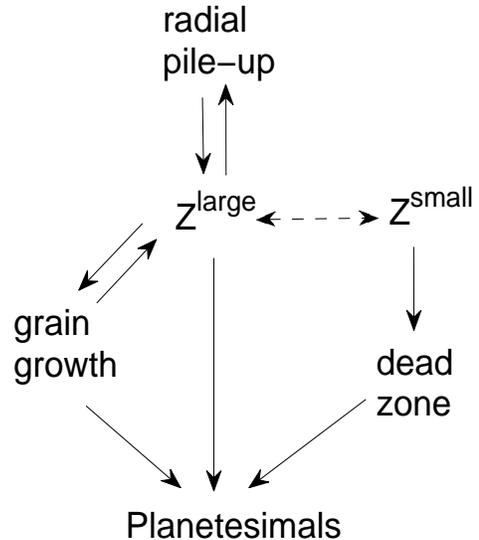}
  \caption{Summary of logical relations between various factors relevant to
  planetesimal formation. A solid arrow from A to B indicates that A promotes
  B. Dashed double arrow indicates speculated connection between the two
  processes. $Z^{\rm large}$ represents metallicity for large particles (millimeter or
  larger), $Z^{\rm small}$ represents the abundance for micron-sized small grains.
  Radial pile-up denotes the pile-up of particles due to the positive radial gradient
  of the radial drift velocity. We refer to ``Grain growth" as collisional
  coagulation towards the largest solid size. ``Dead zone" refers to the radial
  extent of the dead zone. See text for details.
  }\label{fig:summary}
\end{figure}

In this subsection we combine the results summarized in
the previous subsection and discuss various implications for planetesimal
formation. In particular, we emphasize that the importance of the local
enrichment of solid materials in PPDs on planetesimal formation (two feedback
loops, see \S\ref{ssec:loop1} and \S\ref{ssec:loop2}). Our logical chain is
summarized in Figure \ref{fig:summary}, and we elaborate various aspects of this
diagram in the following.

\subsubsection[]{Conditions for Strong Particle Clumping}

Our simulations show a dichotomy in the parameter space in which strong particle
clumping occurs. Strong clumping requires the
presence of relatively large particles with $\tau_s\gtrsim10^{-2}$, and the
abundance of these particles to be super-solar. These two requirements are
represented by the two arrows connecting $Z^{\rm large}$ and
``grain growth" to ``planetesimals" in Figure \ref{fig:summary}. These two
requirements can compensate each other: to form planetesimals, less grain
growth is required if the solid abundance is large enough. In the case that all solids
are strongly coupled to the gas, enhancing the disk metallicity by a factor of 5-30
may be able to trigger GI followed by planetesimal formation
\citep{Sekiya98,YoudinShu02,Chiang08}. 


The particle size that corresponds to $\tau_s=10^{-2}$ depends on the disk model
and distance from the central star. In the MMSN model, this stopping time corresponds
to $2$cm particles at 1AU, and $2$mm or smaller particles at 5AU or beyond,
according to equation (\ref{eq:taus}). These particle sizes are about the maximum
particle size obtained in the recent dust coagulation calculations \citep{Zsom_etal10}.
For more massive disks, the corresponding particle size will be enlarged by a factor
of $f_g$ in the Epstein regime. Since grain growth becomes more difficult when the
particle size exceeds a millimeter, the SI scenario of planetesimal formation prefers
less massive disks, or the outer part of the disk.

\subsubsection[]{Enrichment of Local Solid Abundance}\label{ssec:loop1}

Enrichment of the local abundance of solids is possible by several effects, already
briefly mentioned in \S\ref{sec:intro}. Here we focus on the mechanism proposed
by \citet{YoudinShu02}. For a particle at a fixed size (small, with $\tau_s<1$ at all
disk radii considered), the (single-species) NSH radial drift velocity in PPDs
decreases as the particle drifts inward. Therefore, radial drift causes the
particles to pile-up towards the inner regions, leading to
enhancement of the local abundance of solids. Depending on the time scale, the
largest enhancement factor can reach $3-10$ in $10^5-10^6$ years
\citep{YoudinShu02,YoudinChiang04}, starting in the outer disk and moving
inwards. This process corresponds to the arrow pointing from  ``radial pile-up"
to $Z^{\rm large}$ in Figure \ref{fig:summary}. 

In this paper, we have shown that the radial drift velocity is reduced when local
abundance of solids increases. This effect provides positive feedback to the
enrichment process: particles pile up because the radial drift velocity is smaller at
smaller disk radii. The enrichment of solid material at small disk radii
further reduces the radial drift velocity, leading to stronger pile-ups. This effect
corresponds to the arrow pointing from $Z^{\rm large}$ to ``radial pile-up" in Figure
\ref{fig:summary}. We see that the enrichment of the abundance of solids and the
particle pile-up form a feedback loop that enhance each other. Therefore, we
expect even stronger solid enrichment in the inner region of PPDs than previous
calculations \citep{YoudinShu02,YoudinChiang04} sufficient
for SI and/or GI to form planetesimals within PPD dead zone.

\subsubsection[]{Implications for Grain Growth}\label{ssec:loop2}

The radial drift velocity adopted in grain coagulation models (e.g.,
\citealp{Brauer_etal08a,Birnstiel_etal10,Zsom_etal10}) is generally taken from a
single-species NSH equilibrium. However, we have shown that as particles settle
to disk midplane, the radial drift velocity is reduced due to multi-species effects,
and smallest particles can even drift outward rather than inward. More sophisticated
modeling of grain growth is needed to incorporate the multi-species effects.

One consequence of the multi-species NSH equilibrium is that the enhancement of
local abundance of solids strongly reduces the radial drift velocity, hence the particle
mutual collision velocity. Because the radial drift velocity dominates the collision
velocities for relatively large particles, we expect particles to grow larger in regions
with large abundance of solids, due to the reduced collision velocity there. This effect
is illustrated as an arrow pointing from $Z^{\rm large}$ to ``grain growth" in Figure
\ref{fig:summary}. In turn, grain growth form large particles from smaller ones,
increasing the population of large particles, hence $Z^{\rm large}$, leading to the
arrow pointing in the opposite direction. Again, these two effects form a feedback
loop, promoting grain growth and enrichment of solid material.


\subsubsection[]{Influence from Magnetic Activity}

All our simulations ignore external sources of turbulence, particularly the MRI,
by working in the dead zone of PPDs. Even very weak external turbulence with
$\alpha\sim10^{-4}$ may stir up the solids and maintain them at a height
that may be insufficient for SI to form dense particle clumps (see Table
\ref{tab:vertdiff}). \citet{Johansen_etal07} showed that planetesimal formation
is facilitated by MRI due to the long-lived overdensity regions that effectively
trap the particles. However, they chose very large particle size with
$\tau_s=0.25-1$. When smaller particles are used, they are diffused to a much
larger height. Although SI may still be present with MRI turbulence and form
elongated structures \citep{Balsara_etal09}, particle overdensities are small. Moreover,
the mutual particle collision velocity is much higher in MRI turbulence
\citep{Carballido_etal08}, which inhibits particle growth.
Therefore, we expect the dead zone to be the more favored region for planetesimal
formation, and planetesimal formation should be easier in PPDs with a larger
(radial) extent of the dead zone. This is shown as the arrow pointing from ``dead
zone" towards ``planetesimal formation" in Figure \ref{fig:summary}.

At the same time, the vertical and radial extent of the dead zone in PPDs strongly
depend on the abundance of micron-sized and smaller grains
\citep{IlgnerNelson06,BaiGoodman09}, which is reflected in the arrow connecting
$Z^{\rm small}$ and ``dead zone". Whether there is any connection
between $Z^{\rm large}$ and $Z^{\rm small}$ is uncertain.
Both $Z^{\rm large}$ and $Z^{\rm small}$ should be correlated with the overall disk
metallicity. Moreover, with more large particles (larger $Z^{\rm large}$, collisional
fragmentation may lead to more small grains (larger $Z^{\rm small}$). Despite many
uncertainties, we draw a dashed double-arrow between $Z^{\rm large}$ and
$Z^{\rm small}$ as our speculated connection between the two particle populations.
If such connection exist, it represents the third way for (local) solid abundance
enrichment to promote planetesimal formation.


\subsection[]{Limitations and Outlook}

Our simulations take a local patch from a simple global disk model in which all
physical quantities follow a power-law dependence on disk radii. The global
structure of PPDs may be more complicated. In particular, the presence of a dead
zone in PPDs changes the steady-state disk surface density profile, and may lead
to local pressure maxima at the snow line \citep{KretkeLin07,Brauer_etal08b},
and the inner edge of the dead zone \citep{Dzyurkevich_etal10}. These pressure
bumps are able to trap particles very efficiently. Moreover, solids can also be
trapped in long-lived vortices \citep{BargeSommeria95,KlahrHenning97,
KlahrBodenheimer03,Johansen_etal04,Lyra_etal09}, although their existence
is in debate. Finally, the structure of PPDs may be very non-steady, and
undergo periods of outburst \citep{Zhu_etal09a}. Global models take into
account large-scale variations of disk structure and can follow the disk evolution.
Currently, the numerical resolution required for resolving
the SI is prohibitively high for running 3D global simulations. Nevertheless, one can perform
local simulations in different disk environments, and piece them together to form a
global picture, as in \citet{Tilley_etal10}.

Our simulations focus on the dynamics in the vicinity of the disk
midplane with very limited radial and vertical box sizes and no
magnetic field. This is mainly constrained by the fine grid resolution
required for this study.  In reality, MRI turbulence in the
active layers may excite vertical oscillations in the disk
midplane\citep{FlemingStone03,OshiMacLow09}. Moreover, turbulent diffusion of ions
may produce magnetic activity at the disk midplane, making it
``undead" \citep{TurnerSano08}. Finally, turbulent mixing of particles
may become important when the active layer is relatively thick
\citep{Turner_etal10}. Including these effects into numerical
simulations of planetesimal formation requires enlarging our box
size in all dimensions by a factor of at least 10. Moreover, 3D
rather than 2D simulations are necessary to maintain sustained MRI
turbulence.  Such simulations are computationally expensive, however recently
the static mesh refinement (SMR) algorithm in the Athena MHD code has
been parallelized.  The cost of hybrid (particles and gas)
simulations of layered disks can be substantially reduced by using
fine resolution at the disk midplane (to capture the SI) and in the
active layers (to capture the MRI), while using coarse resolution
everywhere else. With SMR, it becomes feasible to study the effect
of (non-ideal) MRI turbulence in the active layer on particle
dynamics and planetesimal formation in the disk midplane. This is
planned as future work.

\acknowledgments

We thank Eugene Chiang, Jeremy Goodman, Anders Johansen, D.N.C. Lin,
Frank Shu, and Andrew Youdin for useful discussions. This work uses
computational facilities provided by PICSciE at Princeton University, and is
supported by the NSF grant AST-0908269. XNB acknowledges support from
NASA Earth and Space Science Fellowship.

\appendix

\section[]{A. Multi-species NSH equilibrium}\label{app:nsh}

In this appendix we generalize the NSH equilibrium solution to include
multi-species of particles. We start from the force balance for both gas and
particle components. Their velocities are denoted by ${\mb u}$ for gas and
${\mb v}_k$ for particle type $k$. $\tau_k$ and $\epsilon_k$ denote dimensionless
particle stopping time and solid to gas mass ratio for particle type $k$. In
writing down the equations, we subtract both gas and particle velocities by
linear Keplerian shear $-(3/2)\Omega x\hat{\mb{y}}$, and denote the remaining
velocities by $\mb{u}'$ and $\mb{v}'_k$ respectively. Neglecting vertical gravity,
the hydrostatic equilibrium equations read
\begin{subequations}
\begin{align}
&2v'_{ky}\hat{\mb{x}}-\frac{1}{2}v'_{kx}\hat{\mb{y}}-
\frac{1}{\tau_k}(\mb{v}'_k-\mb{u}')=0\ ,\label{eq:app1a}\\
&2u'_{y}\hat{\mb{x}}-\frac{1}{2}
u'_{x}\hat{\mb{y}}+\sum_k\frac{\epsilon_k}{\tau_k}(\mb{v}'_k-\mb{u}')=-2\eta
v_K\hat{\mb{x}}\ .\label{eq:app1b}
\end{align}
\end{subequations}
Multiplying equation (\ref{eq:app1a}) by $\epsilon_k$ for each $k$ and adding
them to equation (\ref{eq:app1b}), we find the expression of the gas velocity in
terms of particle velocities
\begin{equation}
\mb{u}'=-\sum_{k}\epsilon_k\mb{v}'_k-\eta v_K\hat{\mb{y}}\ .
\end{equation}
To obtain the particle velocities, we define velocity vectors
$\Upsilon_x\equiv(v'_{1x}, v'_{2x}, \ldots, v'_{nx})^T$, and
$\Upsilon_y\equiv(v'_{1y}, v'_{2y}, \ldots, v'_{ny})^T$. Further we define a
diagonal matrix $\Lambda\equiv{\rm diag}\{\tau_1, \tau_2, \ldots, \tau_n\}$, and
a rank 1 matrix
$\Gamma\equiv(\mb{\epsilon}, \mb{\epsilon}, \ldots, \mb{\epsilon})^T$, where
$\mb{\epsilon}\equiv(\epsilon_1, \epsilon_2, \ldots, \epsilon_n)^T$. With these
notations, the equations governing $\mb{v}'_k$ can be written as
\begin{equation}\label{eq:equi}
\begin{pmatrix} 1+\Gamma & -2\Lambda \\ \Lambda/2 &
1+\Gamma\end{pmatrix}
\begin{pmatrix} \Upsilon_x\\ \Upsilon_y\end{pmatrix}
=-\eta v_K\begin{pmatrix} 0\\ 1\end{pmatrix}\ .
\end{equation}
The solution of this equation can not be expressed simply; but taking advantage of
the block structure of the  matrices, one can find the solution in the form
\begin{equation}\label{eq:multinsh}
\begin{pmatrix} \Upsilon_x\\ \Upsilon_y\end{pmatrix}=
-\eta v_K\begin{pmatrix} A & 2B \\ -B/2 &
A\end{pmatrix}\begin{pmatrix} 0\\ 1\end{pmatrix}\ ,
\end{equation}
where
\begin{equation}
B=\{[\Lambda^{-1}(1+\Gamma)]^2+1\}^{-1}\Lambda^{-1}\ ,\qquad
A=\Lambda^{-1}(1+\Gamma)B\ .
\end{equation}

One can easily verify that equation (\ref{eq:multinsh}) indeed generalizes the
single species NSH solution. The multi-species solution obtained here is useful
for setting initial conditions of the simulation as well as analysis of
simulation data.

\section[]{B. Collision Velocity Distribution}\label{app:collision}

\begin{figure*}
    \centering
    \includegraphics[width=180mm]{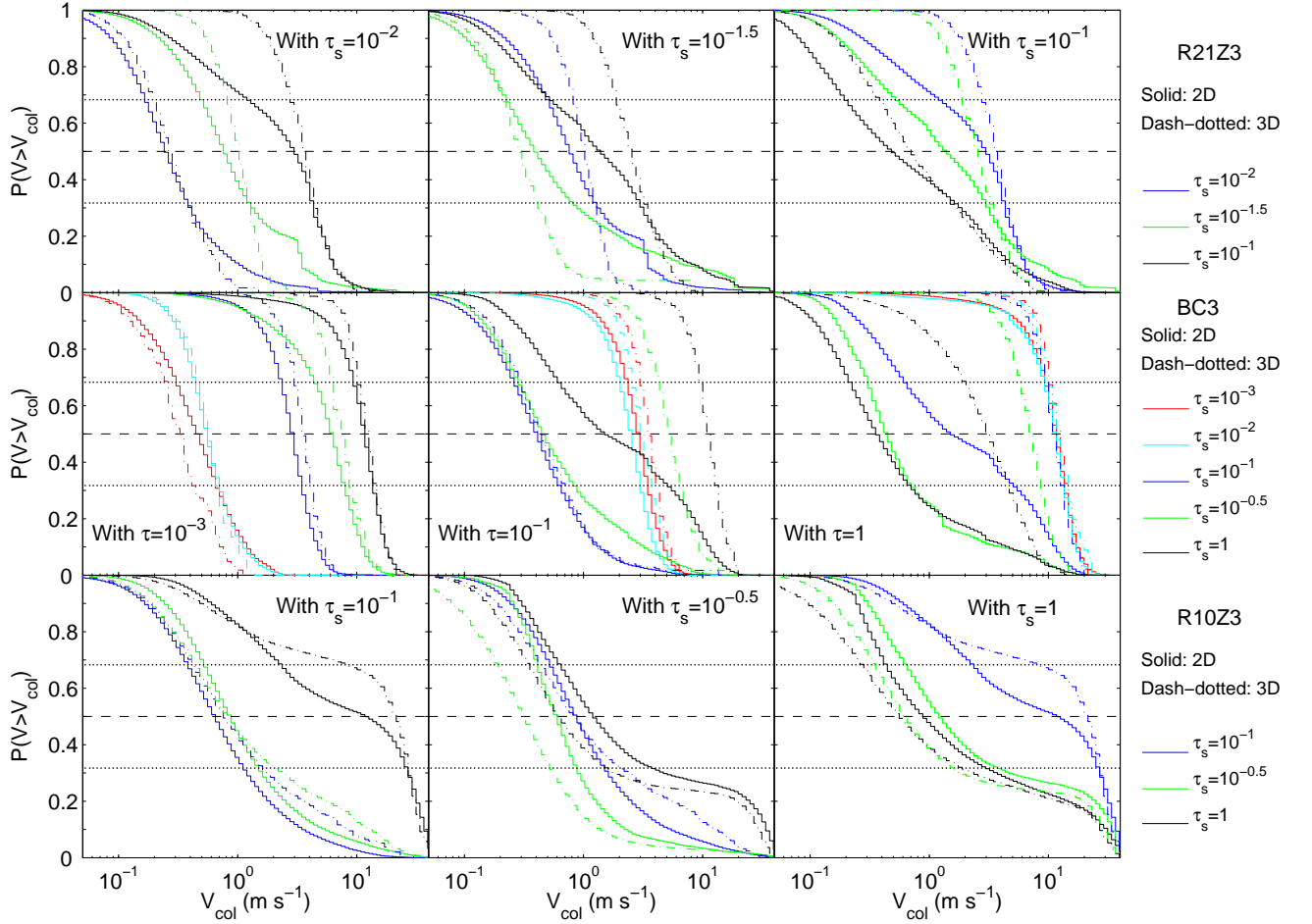}
  \caption{The cumulative probability distribution function (CPDF) of two-body
  particle collision velocities from runs R21Z3 (top), R30Z3
  (middle) and R10Z3 (bottom), illustrating the effect of particle clumping
  on collision velocities. Results from both 2D (solid) and 3D (dash-dotted)
  simulations are shown. In each panel, we plot the CPDFs for collisions between
  various particle species (with different colors, see figure legends) and one
  specific particle species with fixed $\tau_s$ (indicated in each panel).  Dashed
  lines mark the median value, while dotted lines mark the $1\sigma$ level of
  fluctuations. Note that 3D runs in R21 and R30 do not show particle clumping,
  while the others do. }\label{fig:coll_pdf}
\end{figure*}

In this appendix we discuss the distribution of particle collision velocities. In
Figure \ref{fig:coll_pdf} we show the CPDFs from runs R21Z3,
R30Z3 and R10Z3 (for both 2D and 3D). Each panel plots the collision
velocity CPDFs between several (or all) particle species and a given particle
species. The median of the collision velocity is represented by the dashed lines
at $P=0.5$, with $1\sigma$ range obtained by cutting the plots at $P=0.68$ and
$P=0.32$, shown in dotted lines.



Two 3D runs R21Z3-3D and R30Z3-3D do not show strong particle clumping, and
their CPDFs are representative of runs without particle clumping. The CPDF
curves for different pairs of particle types are very similar between each other.
The curves are generally anti-symmetric with respect to the median value $P=0.5$,
and the corresponding velocity distribution is close to log-normal
distribution with super-exponential cutoff at large velocity. Moreover, even in runs
that show strong particle clumping, the CPDFs for collisions between relatively
small particles with $\tau_s\lesssim10^{-2}$ also approaches the log-normal form,
as one can see from run R30Z3-2D. The $1\sigma$ limit of the collision velocity is
generally less than half of the median collision velocity.

The CPDFs from simulations that show strong clumping deviates substantially
from log-normal. For runs R21Z3 and R30Z3, the
collision velocity is clearly reduced in 2D relative to 3D
(2D runs show clumping while 3D runs do not). In run R30Z3-2D, we
see that the reduction is most significant for collisions between two relatively large
particles with $\tau_s\gtrsim0.1$, which are the ones that make up most of the
clumps (see Figure \ref{fig:dparpdf}). The collision velocity between large
($\tau_s>0.1$) and small ($\tau_s<0.1$) particles is also reduced, although to a
lesser extent. For R21Z3-2D and R10Z3 (both 2D and 3D) runs, all particles actively
participate in the SI and clumping,  and their mutual collision velocities are all reduced.
The reduction of collision velocity appears to be stronger in R30Z3 and R10Z3
runs, where $\tau_{\rm max}=1$. Finally, we see that the reduction of collision
velocity is most significant for intermediate high-velocity collisions
($V_{\rm col}\simeq1-5$m s$^{-1}$), although there is still a high-velocity tail
present. It is likely that the high-velocity tail is caused by collisions outside
the particle clump.

\begin{figure*}
    \centering
    \includegraphics[width=150mm,height=125mm]{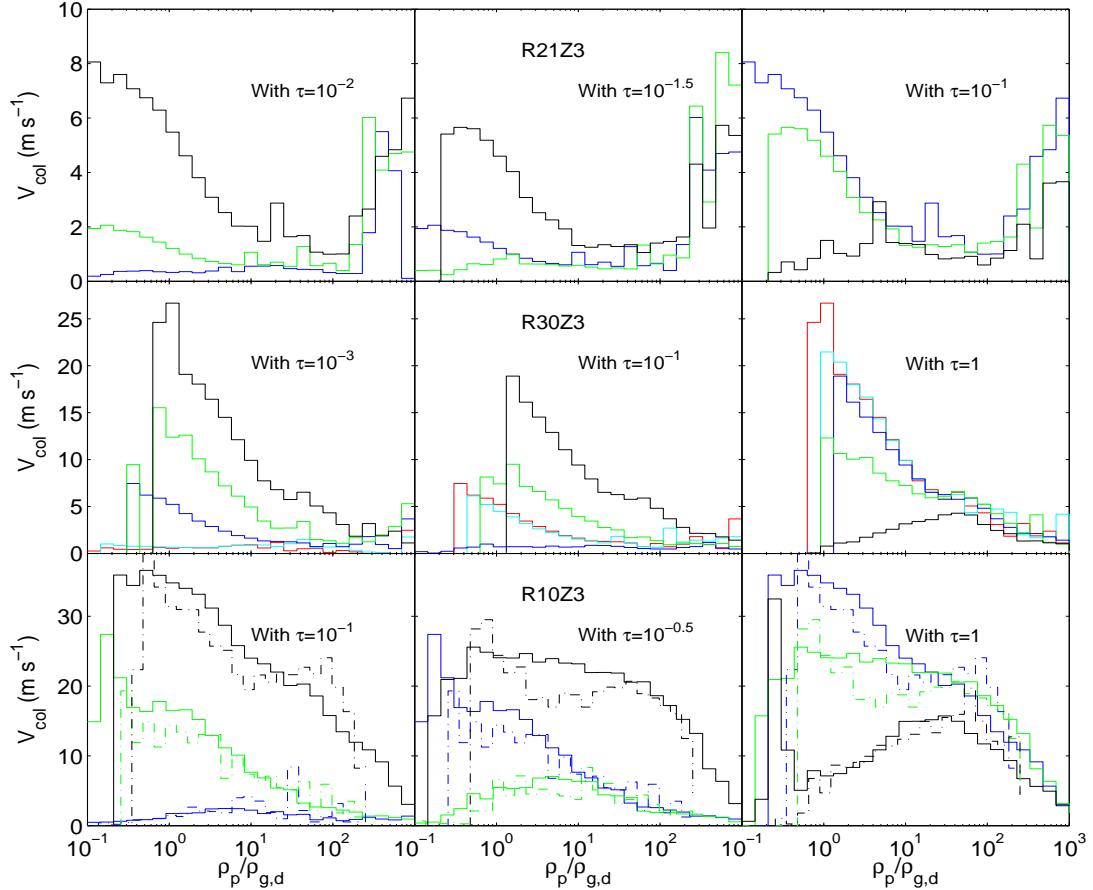}
  \caption{The mean particle collision velocity as a function of particle density
  for the three runs that show particle clumping (R21Z3, R30Z3 and
  R10Z3, from top to bottom). The 2D results are shown in solid lines, while 3D
  results are shown as dash-dotted lines (bottom panels only). In each
  panel, we show the median collision velocity between various
  particle species (using the same color scheme in Figure
  \ref{fig:coll_pdf}) and a specific particle species with a $\tau_s$ as
  indicated in each panel.}\label{fig:coll_dpar}
\end{figure*}

To further address the influence of clumping on particle collision velocities, we
evaluate the mean collision velocity as a function of ambient particle density
$\rho_p$. For the three sets of runs that show particle clumping R21Z3 (2D),
R30Z3 (2D) and R103 (2D and 3D), we show the results in Figure
\ref{fig:coll_dpar}. For both R30Z3 and R10Z3 runs, we find a clear trend that
the collision velocity is strongly reduced towards higher $\rho_p$. More
specifically, the reduction is most prominent for collisions between large and
small particles. The collision velocity between particles of similar sizes appears
to be insensitive to $\rho_p$, and is maintained at a relatively low value.
Interestingly, results from run R21Z3-2D show a different behavior. The collision
velocity decreases with $\rho_p$ until $\rho_p/\rho_{g,b}\sim300$. Beyond this
density, we observe an increase of collision velocity towards larger $\rho_p$.
We have performed an additional R41 run in 2D with Z=0.05, which shows
particle clumping, and the collision velocity shows similar properties as in
R21Z3-2D. This is very likely to be due to the different properties of the SI
for particles with $\tau_s=0.1$ from particles with $\tau_s=1$. As shown in
non-stratified simulations \citep{JohansenYoudin07}, clumps are much more
dynamic in the former case, which may lead to larger collision velocities.


\label{lastpage}
\end{document}